\pdfoutput=1
\documentclass[11pt]{article}

\usepackage[final]{acl}

\usepackage{times}
\usepackage{latexsym}
\usepackage{pdfpages}
\usepackage[T1]{fontenc}

\usepackage[utf8]{inputenc}

\usepackage{microtype}

\usepackage{inconsolata}
\usepackage{enumitem} 
\usepackage{float}

\usepackage{graphicx}
\usepackage{microtype}

\usepackage{subfigure}
\usepackage{booktabs} 
\usepackage{tabularx} 
\usepackage{multirow} 
\usepackage{hyperref}
\usepackage{listings}  
\usepackage{color}     
\usepackage{adjustbox} 
\usepackage{amsmath}
\usepackage{amssymb}
\usepackage{mathtools}
\usepackage{amsthm}
\usepackage[capitalize,noabbrev]{cleveref}

\usepackage{tikz}
\usetikzlibrary{shapes.geometric, arrows, positioning, fit, backgrounds, matrix}

\title{LibVulnWatch: A Deep Assessment Agent System and Leaderboard for Uncovering Hidden Vulnerabilities in Open-Source AI Libraries}

\author{
\begin{tabular}{c}
Zekun Wu\textsuperscript{1,2}\thanks{Equal contributions} \quad
Seonglae Cho\textsuperscript{1,2}\footnotemark[1] \quad
Umar Mohammed\textsuperscript{1} \quad
Cristian Munoz\textsuperscript{1} \\[4pt]
Kleyton Costa\textsuperscript{1} \quad
Xin Guan\textsuperscript{1} \quad
Theo King\textsuperscript{1} \quad
Ze Wang\textsuperscript{1,2} \\[4pt]
Emre Kazim\textsuperscript{1,2} \quad
Adriano Koshiyama\textsuperscript{1,2}\thanks{Corresponding author} \\[6pt]
\textsuperscript{1}Holistic AI \quad
\textsuperscript{2}University College London
\end{tabular}
}

\begin{document}
\maketitle
\begin{abstract}
Open-source AI libraries are foundational to modern AI systems, yet they present significant, underexamined risks spanning security, licensing, maintenance, supply chain integrity, and regulatory compliance. We introduce \textsc{LibVulnWatch}, a system that leverages recent advances in large language models and agentic workflows to perform deep, evidence-based evaluations of these libraries. Built on a graph-based orchestration of specialized agents, the framework extracts, verifies, and quantifies risk using information from repositories, documentation, and vulnerability databases. \textsc{LibVulnWatch} produces reproducible, governance-aligned scores across five critical domains, publishing results to a public leaderboard for ongoing ecosystem monitoring. Applied to 20 widely used libraries—including ML frameworks, LLM inference engines, and agent orchestration tools—our approach covers up to 88\% of OpenSSF Scorecard checks while surfacing up to 19 additional risks per library, such as critical RCE vulnerabilities, missing SBOMs, and regulatory gaps. By integrating advanced language technologies with the practical demands of software risk assessment, this work demonstrates a scalable, transparent mechanism for continuous supply chain evaluation and informed library selection.
\end{abstract} 

\section{Introduction}
\label{sec:introduction}

The rapid adoption of AI systems in high-stakes domains has intensified the need for robust technical governance and risk assessment. While policy frameworks increasingly call for transparency, accountability, and safety, a persistent gap remains between these governance objectives and the engineering practices required to realize them~\cite{reuel2025open}. Open-source libraries and frameworks, which underpin most modern machine learning systems, introduce complex legal, security, maintenance, and regulatory risks that are often overlooked by conventional assessment tools~\cite{Wang2025,Alevizos_2024}. These tools typically provide surface-level checks and lack the depth needed to uncover nuanced vulnerabilities in the AI software supply chain.

Recent progress in large language models and agentic workflows has enabled new approaches to structured, evidence-based analysis across diverse domains. In this work, we introduce \textsc{LibVulnWatch}, a system that leverages these advances to perform deep, multi-domain evaluations of open-source AI libraries. The system coordinates specialized agents to assess five critical risk domains—licensing, security, maintenance, dependency management, and regulatory compliance—drawing on verifiable evidence from repositories, advisories, and documentation.

To enable continuous ecosystem monitoring and evidence-based decision-making, we publish every assessment on a public leaderboard\footnote{The leaderboard and all per-library assessment reports are publicly available on Hugging Face at \url{https://huggingface.co/spaces/holistic-ai/LibVulnWatch}.}. Evaluating 20 widely used AI libraries—including ML frameworks, inference engines, and agent orchestration tools—\textsc{LibVulnWatch} demonstrates:
\begin{itemize}[nosep]
    \item Up to \textbf{88\% coverage} of OpenSSF Scorecard checks;
    \item \textbf{Up to 19 additional risks} per library, including RCEs, missing SBOMs, and compliance gaps;
    \item \textbf{Governance-aligned, reproducible scores} for transparent comparison and risk management.
\end{itemize}

By integrating advanced language technologies with the practical demands of software risk assessment, \textsc{LibVulnWatch} offers a scalable, transparent mechanism for operationalizing governance principles in open-source AI infrastructure.
\section{Related Work}

Research on vulnerabilities in AI pipelines has expanded beyond adversarial inputs and data poisoning to encompass system-level risks in the software supply chain~\cite{Wang2025}. Studies have analyzed large-scale LLM supply chain issues, revealing flaws in application and serving components, while others have documented recurring bugs in widely used frameworks such as TensorFlow and PyTorch~\cite{Chen2023}. LLM-based vulnerability detection has shown promise for code analysis~\cite{Zhou2024}, though challenges such as false positives and domain adaptation remain. Broader supply chain threats—including dependency confusion and package hijacking—are well-documented~\cite{Ladisa2023,Ohm2020}.

Efforts to assess open-source project hygiene, such as the OpenSSF Scorecard~\cite{Zahan2023}, provide valuable surface metrics but often lack the depth required for comprehensive vulnerability analysis. Recent advances in multi-agent orchestration frameworks, including LangChain and LangGraph~\cite{LangChain2025,LangGraph2025}, have enabled more structured and scalable approaches to information extraction and evaluation, forming the basis for several assessment pipelines.
\section{Methodology}
\label{sec:methodology}

Our approach leverages recent advances in language models and multi-agent systems to address complex challenges in software risk assessment. By adapting NLP techniques for information extraction, knowledge synthesis, and structured reasoning, we operationalize key Technical AI Governance capacities through a multi-stage evaluation pipeline. This section details the pipeline's architecture, risk assessment framework, evaluation protocol, and benchmarking procedures.

\subsection{Risk Assessment Framework}
\label{subsec:risk-framework}

We define a comprehensive risk assessment framework adapted from established open-source and AI risk taxonomies. It encompasses five governance-relevant domains, each with specific factors for evaluation:
\begin{itemize}[nosep]
    \item \textbf{License Analysis:} Assessing license type (e.g., MIT, Apache 2.0, GPL), version, commercial use compatibility, distribution rights, patent grant provisions, attribution requirements, and overall conformance with open-source compliance standards.
    \item \textbf{Security Assessment:} Evaluating known Common Vulnerabilities and Exposures (CVEs) within the last 24 months (count and severity), the existence and adequacy of a security disclosure policy, responsiveness to security issues, evidence of security testing (e.g., CI/CD test coverage), and the handling of released binaries or signed artifacts.
    \item \textbf{Maintenance Indicators:} Analyzing release frequency and the date of the latest release, the number and activity levels of contributors (including diversity and organizational backing), issue resolution metrics (e.g., response times, recent commit activity), and the project governance model and packaging workflow.
    \item \textbf{Dependency Management:} Examining Software Bill of Materials (SBOM) availability and format (e.g., CycloneDX, SPDX), direct and transitive dependency counts, policies and tools for dependency updates, and the identification of known vulnerable dependencies.
    \item \textbf{Regulatory Considerations:} Reviewing documentation for alignment with relevant compliance frameworks (e.g., GDPR, AI Act), the availability of explainability features (especially for AI/ML libraries), stated data privacy provisions, and the presence of audit documentation or support for audit readiness.
\end{itemize}
Each of these five domains, as depicted as parallel tracks at the top of Figure~\ref{fig:assessment-architecture}, is operationalized as a distinct assessment module within the agentic workflow, guided by engineered prompts enforcing key concept coverage and quantifiable metric extraction.

\subsection{Agentic Workflow}
\label{subsec:agentic-graph}

Our system employs a structured, agentic workflow implemented as a DAG using a modern agent orchestration framework. Our implementation was inspired by the Open Deep Research repository\footnote{\url{https://github.com/langchain-ai/open_deep_research}}. We redesigned the graph design and defined domain-specific prompts that adapt language model capabilities to the specific knowledge requirements of security, licensing, and compliance assessment. All experiments used \texttt{gpt-4.1-mini} (costing approx. \$0.10 per library). OpenSSF Scorecard \citep{Zahan2023} checks were run on the primary GitHub repository of each target library, and we used the Google Search API for evidence retrieval. 

\begin{figure}[t!]
    \centering
    \includegraphics[width=0.47\textwidth]{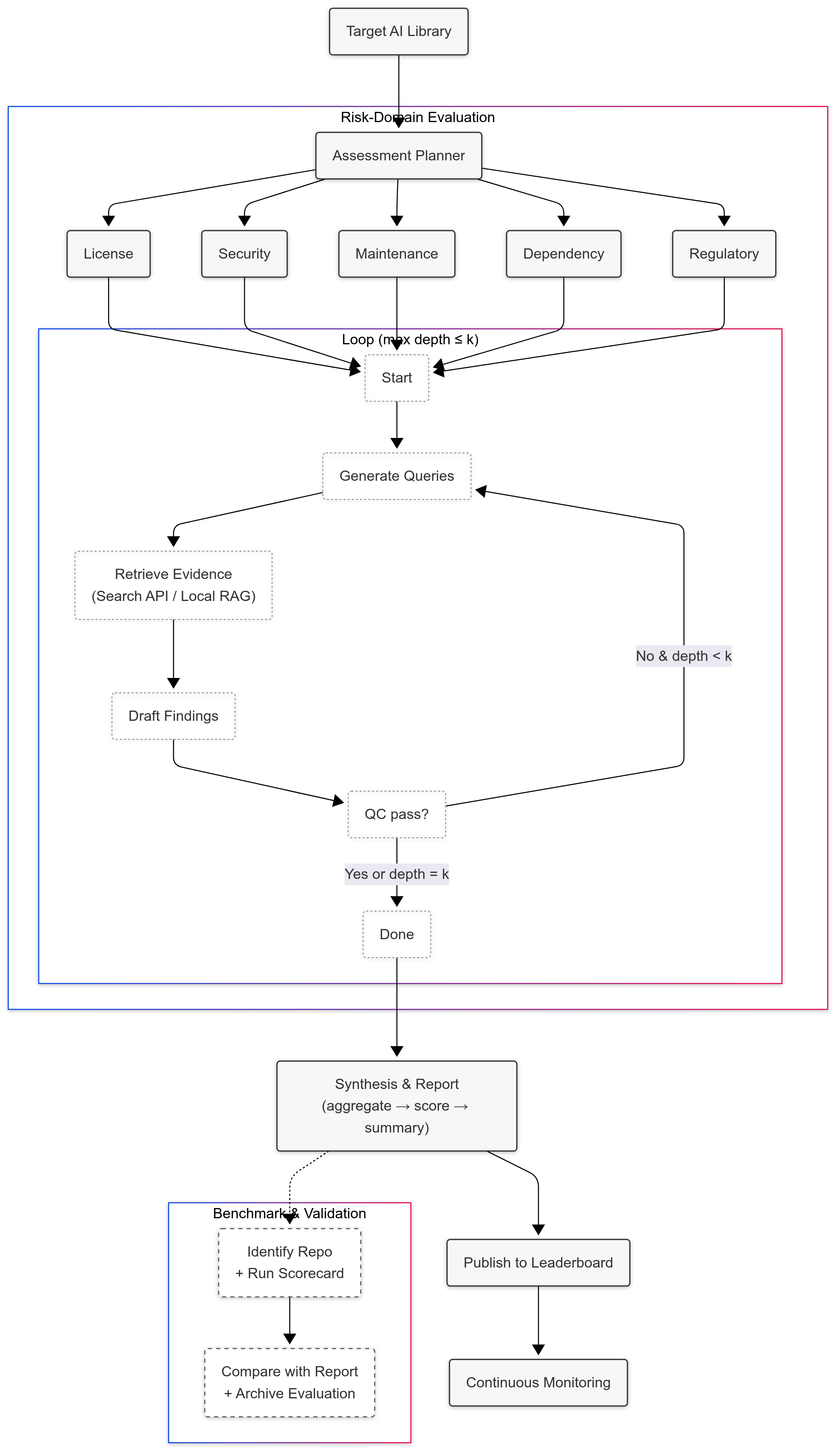}
    \caption{Workflow of the automated agent. Each risk domain (License, Security, Maintenance, Dependency, Regulatory) runs in parallel, with controlled-depth evidence retrieval and drafting. The results are synthesized into a report, benchmarked using the OpenSSF Scorecard, and then published with monitoring.}
    \label{fig:assessment-architecture}
\end{figure}

The automated workflow addresses particular challenges of applying language models to evidence-based assessment, including factuality verification and domain-specific knowledge extraction. It begins with high-level search-based planning, followed by domain-specific iterative retrieval until sufficient evidence is gathered for each of the five domains. These are processed in parallel to generate draft findings, which are combined into a full report including an executive summary. The report is then validated by identifying the main GitHub repository, running the Scorecard, and comparing outputs using an LLM. This approach ensures modularity, consistency, and parallelism. Integrating the LLM's text understanding, structured data handling, and search capabilities, the overall agentic workflow is illustrated in Figure~\ref{fig:assessment-architecture} and comprises the following key stages:
\begin{itemize}[nosep]
    \item \textbf{Planning:} An initial \textit{Assessment Planner} agent (top of Figure~\ref{fig:assessment-architecture}) generates a detailed assessment plan for the target library, adhering strictly to the five core risk domains detailed in Section~\ref{subsec:risk-framework} and formulates initial research queries.
    \item \textbf{Iterative Evidence Gathering and Drafting (Per Domain):} For each of the five risk domains, operating in parallel:
    \begin{itemize}[nosep]
        \item \textit{Query Generation:} Targeted search queries are formulated.
        \item \textit{Evidence Retrieval:} A dedicated agent iteratively performs searches against authoritative sources (e.g., official documentation, security databases, repository metadata using specialized query patterns via Search API / Local RAG) to aggregate evidence. This includes the use of advanced search operators and repository-specific query patterns (e.g., for GitHub) to extract structured data and metrics where direct API access is not assumed.
        \item \textit{Draft Findings:} The retrieved evidence is synthesized into initial draft findings for the specific domain.
        \item \textit{Quality Check \& Refinement Loop:} A quality check (QC) assesses if sufficient evidence has been gathered and if the findings meet predefined criteria. If the QC is not passed and the maximum search depth (k) has not been reached, the process loops back to generate refined queries and retrieve more evidence. This iterative loop continues until the QC is passed or the depth limit is reached.
    \end{itemize}
    This entire synthesis process is strictly governed by prompts engineered to adapt language understanding capabilities to the software security context, enforcing structured reporting (e.g., with sections for an executive overview, emergency issues, and a detailed table of findings with columns for Risk Factor, Observed Data, Rating, Reason for Rating, and Key Control), quantification, evidence citation, and handling of missing information. The specific instruction sets (prompts) used for each key agent are detailed in Appendix~\ref{subsec:AgentPrompts}.
    \item \textbf{Synthesis \& Report Compilation:} Once drafting for all domains is complete (marked as Done in Figure~\ref{fig:assessment-architecture}), a final agent synthesizes the individual domain findings into a consolidated, structured report. This includes an executive summary, a risk dashboard, highlighted emergency issues, prioritized controls, and a mitigation strategy.
    \item \textbf{Benchmark Validation:} Before final publication, the generated report undergoes a validation step. This involves identifying the main repository of the target library, running the OpenSSF Scorecard, and comparing the Scorecard output with the agent\'s report (often using an LLM for an \textit{Archive Evaluation}) to assess alignment and novelty, as depicted in Figure~\ref{fig:assessment-architecture}.
    \item \textbf{Public Reporting and Ecosystem Monitoring:} The validated and finalized report is programmatically published to a public leaderboard, which is implemented as an interactive Gradio application hosted on Hugging Face Spaces (see Appendix~\ref{subsec:LeaderboardInterface} for details and screenshots). This facilitates \textit{Ecosystem Monitoring} and accountability by dynamically ranking libraries by Trust Score and highlighting key risks. We follow responsible disclosure practices for any new, non-public vulnerabilities identified during the assessment.
\end{itemize}

\subsection{Evaluated AI Libraries}
\label{subsec:evaluated-libraries}

We evaluated 20 diverse open-source AI libraries spanning the AI lifecycle, selected for representative coverage (see Table~\ref{tab:coverage-summary} for list and scores). Libraries were chosen from three key functional categories, aiming for diversity in function, community size, maturity, and impact:
\begin{itemize}[nosep]
    \item \textbf{Core ML/DL Frameworks:} PyTorch \citep{PaszkeGMLBCKLGA19}, TensorFlow \citep{AbadiBCCDDDGIIK16}, ONNX \citep{ONNX2025}, Huggingface Transformers \citep{WolfDSCDMCRLFDS20}, and JAX \citep{bradbury2018jax}.
    \item \textbf{LLM Inference \& Orchestration Tools:} TensorRT \citep{TensorRT2025}, LlamaIndex \citep{LiuLlamaIndex2022}, SGLang \citep{zheng2023sglang}, vLLM \citep{kwon2023efficient}, LangChain \citep{LangChain2025}, and Text Generation Inference \citep{TextGenInference2025}.
    \item \textbf{AI Agent Frameworks:} Browser Use \citep{browseruse2024}, CrewAI \citep{CrewAI2025}, MetaGPT \citep{zhang2024metagpt}, LangGraph \citep{LangGraph2025}, SmolAgents \citep{smolagents}, Stagehand \citep{Stagehand2025}, Composio \citep{Composio2025}, Pydantic AI \citep{PydanticAI2025}, and Agent Development Kit \citep{ADK2025}.
\end{itemize}
Each library underwent the full protocol; results are public.

\subsection{Risk Scoring}
\label{subsec:quantification}

We employ a 1-5 numerical scale for risk rating within each of the five governance-relevant domains outlined above (Section~\ref{subsec:risk-framework}), where 1 indicates High Risk, 3 Medium Risk, and 5 Low Risk. As detailed in the workflow description (Section~\ref{subsec:agentic-graph}), each rating requires justification tied to specific, verifiable evidence thresholds defined in the prompts. \textbf{The risk scoring within each domain is anchored by the following criteria derived from the agent system prompts:}

\begin{itemize}[nosep]
    \item \textbf{Low Risk (Score 5)} is indicated by: \textit{License:} Permissive (e.g., MIT, Apache 2.0, BSD) with clear terms and compatibility; \textit{Security:} No CVEs in the past 24 months, a robust security policy, and rapid fixes (e.g., $<$7 days); \textit{Maintenance:} More than 10 active contributors, monthly or more frequent releases, and prompt issue response (e.g., $<$24 hours); \textit{Dependencies:} SBOM available, fewer than 20 direct dependencies, and evidence of automatic updates; \textit{Regulatory:} Clear compliance documentation and a complete audit trail.
    \item \textbf{Medium Risk (Score 3)} is indicated by: \textit{License:} Moderate restrictions or unclear patent provisions; \textit{Security:} 1-3 minor CVEs in the past 12 months, a basic security policy, and moderate response times (e.g., 7-30 days); \textit{Maintenance:} 3-10 active contributors, quarterly releases, and issue response times of 1-7 days; \textit{Dependencies:} Partial SBOM, 20-50 direct dependencies, and some transitive visibility; \textit{Regulatory:} Incomplete compliance documentation or partial audit readiness.
    \item \textbf{High Risk (Score 1)} is indicated by: \textit{License:} Restrictive terms (e.g., GPL/AGPL), incompatible terms, or other legal concerns; \textit{Security:} Critical or multiple CVEs, a missing security policy, or slow response times (e.g., $>$30 days); \textit{Maintenance:} Fewer than 3 active contributors, infrequent releases (e.g., $>$6 months), or poor issue response; \textit{Dependencies:} No SBOM, more than 50 direct dependencies, or known vulnerable transitive dependencies; \textit{Regulatory:} Missing compliance documentation or failure to meet essential regulations.
\end{itemize}

Critically, the absence of necessary information for assessment (e.g., no public security policy or SBOM) on any key risk factor is also explicitly defined as a High Risk indicator (Score 1). Furthermore, the system is designed to critically evaluate all available information to identify the most significant or concerning risk factor within each domain, even if other factors appear satisfactory, ensuring a thorough and conservative risk posture.
Intermediate scores (2 or 4) may be assigned based on the agent's assessment when evidence suggests a risk level between these defined thresholds. The overall Trust Score provides a composite measure by aggregating the five domain scores ($Li, Se, Ma, De, Re$): 
$\text{Trust}(l) = \frac{1}{5}\sum_{d \in \{Li, Se, Ma, De, Re\}} d(l)$.

\subsection{Benchmarking and Novelty Analysis}
\label{subsec:baseline-metrics}

We use the OpenSSF Scorecard~\citep{Zahan2023} as a baseline to evaluate our agent. This involves identifying the main repository, running the Scorecard, and comparing its output with our agent\'s report to derive two key metrics:

\begin{itemize}[nosep]
    \item \textbf{Baseline Alignment(\%)}: The percentage of relevant Scorecard checks addressed in the agent\'s report, calculated against applicable checks (i.e., excluding checks with non-conclusive scores such as \'?\') from the Scorecard output. This is calculated as $\text{Coverage} = \frac{\text{\# matched checks}}{\text{\# applicable checks}} \times 100$.
    
    \item \textbf{Novelty Yield (\#)}: The number of unique, meaningful issues or deeper contextual insights identified by the agent but not explicitly surfaced by the Scorecard. This is defined as $\text{Yield} = \text{\# unique agent-only findings}$.
\end{itemize}

\section{Results}
\label{sec:results}

Our methodology identified novel vulnerabilities in Open-source AI libraries, often missed by static analysis. Benchmarking against OpenSSF Scorecard~\citep{Zahan2023}, detailed in Section~\ref{subsec:benchmark-analysis}, quantified alignment and unique contextual findings. Section~\ref{subsec:case-studies} presents illustrative examples. For a detailed example of a full assessment output (the analysis report) for the JAX library, please see Appendix~\ref{subsec:ExampleJAXAssessment}; its corresponding baseline evaluation is presented in Appendix~\ref{subsec:JAXReportEvaluation}.

\subsection{Benchmarking and Alignment Analysis}
\label{subsec:benchmark-analysis}

We benchmarked our agentic system against the OpenSSF Scorecard to evaluate alignment and identify unique contributions. Table~\ref{tab:coverage-summary} presents key metrics defined in Section~\ref{sec:methodology}—Baseline Alignment (overlap with Scorecard checks) and Novelty Yield (unique findings)—across all evaluated libraries, grouped by functional category and including category averages. While observed Baseline Alignment for most libraries ranged from 55\% to 88\%, indicating substantial overlap, the agentic system consistently surfaced a significant Novelty Yield (typically 5-13 unique findings per library) not captured by baseline tools. 

The agents showed particular strengths in connecting disparate information sources and contextualizing findings, though they sometimes missed formal contributor declarations, CI testing evidence, binary artifact identification, and explicit security testing policies flagged by the baseline. This suggests opportunities for complementary approaches combining structured checks with context-aware reasoning. Examples of critical risks identified through contextual analysis that went beyond conventional automated scans, contributing to Novelty Yield, include:
\begin{itemize}[nosep]
    \item Complex RCEs from insecure defaults or subtle data processing flaws.
    \item Systemic SBOM absence and supply chain/transitive dependency risks.
    \item Pervasive regulatory/privacy compliance gaps (GDPR, HIPAA, AI Act).
    \item Widespread lack of governance mechanisms (audit trails, explainability, privacy controls).
    \item Undocumented telemetry/data collection (e.g., in one AI agent framework).
    \item Potential patent risks from unclear/insufficient licensing for core ML algorithms.
\end{itemize}

\begin{table}[ht]
\centering
\caption{Baseline Alignment and Novelty Yield Across Libraries}
\label{tab:coverage-summary}
\small
\begin{tabular}{@{}p{3.5cm}cc@{}}
\toprule
\textbf{Library} & \shortstack{\textbf{Baseline} \\ \textbf{Alignment (\%)}} & \shortstack{\textbf{Novelty} \\ \textbf{Yield (\#)}} \\
\midrule
\multicolumn{3}{@{}l}{\textit{Core ML/DL Frameworks} \hfill \textit{\textbf{77.1}} \hfill \hspace{0.5em}\textit{6.8}}\hspace{0.3em} \\
\midrule
PyTorch & 88.2 & 8 \\
JAX & 61.1 & 12 \\
Tensorflow & 72.2 & 5 \\
ONNX & 87.5 & 5 \\
Huggingface Transformers & 76.5 & 4 \\
\midrule
\multicolumn{3}{@{}l}{\textit{LLM Inference \& Orchestration} \hfill \textit{73.7} \hfill \hspace{3em}\textit{7.8}}\hspace{0.3em} \\
\midrule
TensorRT & 68.8 & 5 \\
LlamaIndex & 82.4 & 7 \\
SGLang & 73.3 & 5 \\
vLLM & 73.3 & 7 \\
LangChain & 72.2 & 19 \\
Text Generation Inference & 72.2 & 6 \\
\midrule
\multicolumn{3}{@{}l}{\textit{AI Agent Frameworks} \hfill \quad\quad\textit{76.2} \hfill \hspace{1em}\textit{\textbf{9.1}}}\hspace{0.3em} \\
\midrule
Browser Use & 88.2 & 7 \\
CrewAI & 71.4 & 13 \\
MetaGPT & 57.1 & 7 \\
LangGraph & 77.8 & 7 \\
SmolAgents & 73.3 & 9 \\
Stagehand & 83.3 & 6 \\
Composio & 68.8 & 5 \\
Pydantic AI & 88.2 & 10 \\
Agent Development Kit & 77.9 & 7 \\
\bottomrule
\end{tabular}
\end{table}

\subsection{Aggregated Domain Risk Findings and Patterns}
\label{subsec:domain-findings-patterns}

Table~\ref{tab:risk-assessment-scores-main} presents the detailed library-by-library trust scores across the five primary domains and the composite Trust Score. The context-sensitive analysis enabled by our approach revealed nuanced patterns across evaluated libraries that would be difficult to detect with traditional rules-based assessment. Aggregate Trust Scores varied by category, with Core ML/DL frameworks generally scoring higher than newer AI Agent frameworks, potentially reflecting greater maturity. Common weaknesses were observed across the ecosystem, particularly in:
\begin{itemize}[nosep]
    \item \textbf{Dependency Management:} Widespread absence of SBOMs hindering transparency, poorly managed transitive dependencies, and lack of automated vulnerability scanning were common.
    \item \textbf{Regulatory Considerations:} Significant gaps existed regarding comprehensive documentation for GDPR/HIPAA/AI Act compliance and features for model explainability or audit logging.
    \item \textbf{Security:} Many libraries exhibited vulnerabilities like RCEs, unsigned releases, and insecure CI/CD pipelines, with newer frameworks often lacking mature disclosure policies.
    \item \textbf{License Analysis:} While often permissive, nuanced risks like potential patent issues or conflicts with restrictive licenses (e.g., AGPL) were found, and formal patent grants were frequently missing.
    \item \textbf{Maintenance Indicators:} Established libraries showed robust core maintenance, but patterns of unmaintained sub-projects or less transparency/slower resolution in newer frameworks posed risks.
\end{itemize}

\begin{table}[t!]
\centering
\caption{Detailed Risk Assessment Scores Across Libraries and Domains (Li: License, Se: Security, Ma: Maintenance, De: Dependency, Re: Regulatory, Trust: Trust Score; Scale: 1-5, higher is better)}
\label{tab:risk-assessment-scores-main}
\footnotesize
\begin{adjustbox}{width=\columnwidth,center}
\begin{tabular}{@{}p{3.2cm}cccccc@{}}
\toprule
\textbf{Library} & \textbf{Li} & \textbf{Se} & \textbf{Ma} & \textbf{De} & \textbf{Re} & \textbf{Trust} \\
\midrule
\multicolumn{7}{@{}l}{\textit{Core ML/DL Frameworks} \hfill \textit{13.0}}\hspace{-0.6em} \\
\midrule
PyTorch & 5 & 1 & 3 & 1 & 3 & 13 \\
JAX & 5 & 3 & 4 & 1 & 1 & \textbf{14} \\
Tensorflow & 5 & 1 & 3 & 1 & 3 & 13 \\
ONNX & 5 & 1 & 3 & 1 & 1 & 11 \\
Transformers & 5 & 1 & 4 & 1 & 3 & \textbf{14} \\
\midrule
\multicolumn{7}{@{}l}{\textit{LLM Inference \& Orchestration} \hfill \textit{11.8}}\hspace{-0.6em} \\
\midrule
TensorRT & 5 & 1 & 5 & 1 & 3 & \textbf{15} \\
LlamaIndex & 5 & 1 & 3 & 1 & 3 & 13 \\
SGLang & 5 & 1 & 3 & 1 & 1 & 11 \\
vLLM & 3 & 1 & 4 & 1 & 1 & 10 \\
LangChain & 5 & 1 & 1 & 1 & 3 & 11 \\
Text Generation Inference & 5 & 1 & 3 & 1 & 1 & 11 \\
\midrule
\multicolumn{7}{@{}l}{\textit{AI Agent Frameworks} \hfill \textit{11.4}}\hspace{-0.6em} \\
\midrule
CrewAI & 5 & 1 & 3 & 1 & 1 & 11 \\
MetaGPT & 5 & 1 & 5 & 1 & 1 & 13 \\
LangGraph & 1 & 1 & 3 & 1 & 3 & 9 \\
SmolAgents & 5 & 1 & 1 & 1 & 1 & 9 \\
Stagehand & 5 & 3 & 1 & 1 & 1 & 11 \\
Composio & 1 & 1 & 5 & 1 & 3 & 11 \\
Browser Use & 5 & 1 & 4 & 1 & 3 & \textbf{14} \\
Pydantic AI & 5 & 1 & 3 & 1 & 1 & 11 \\
Agent Development Kit & 5 & 3 & 4 & 1 & 1 & \textbf{14} \\
\bottomrule
\end{tabular}
\end{adjustbox}
\end{table}

\subsection{Illustrative Case Studies}
\label{subsec:case-studies}

To further illustrate the capabilities of \textsc{LibVulnWatch}, we present five case studies highlighting how semantic understanding and contextual analysis revealed insights that would be challenging to capture through traditional assessment approaches.

\subsubsection*{License Analysis: LangGraph}
\begin{figure}[h]
    \centering
    \includegraphics[width=\linewidth]{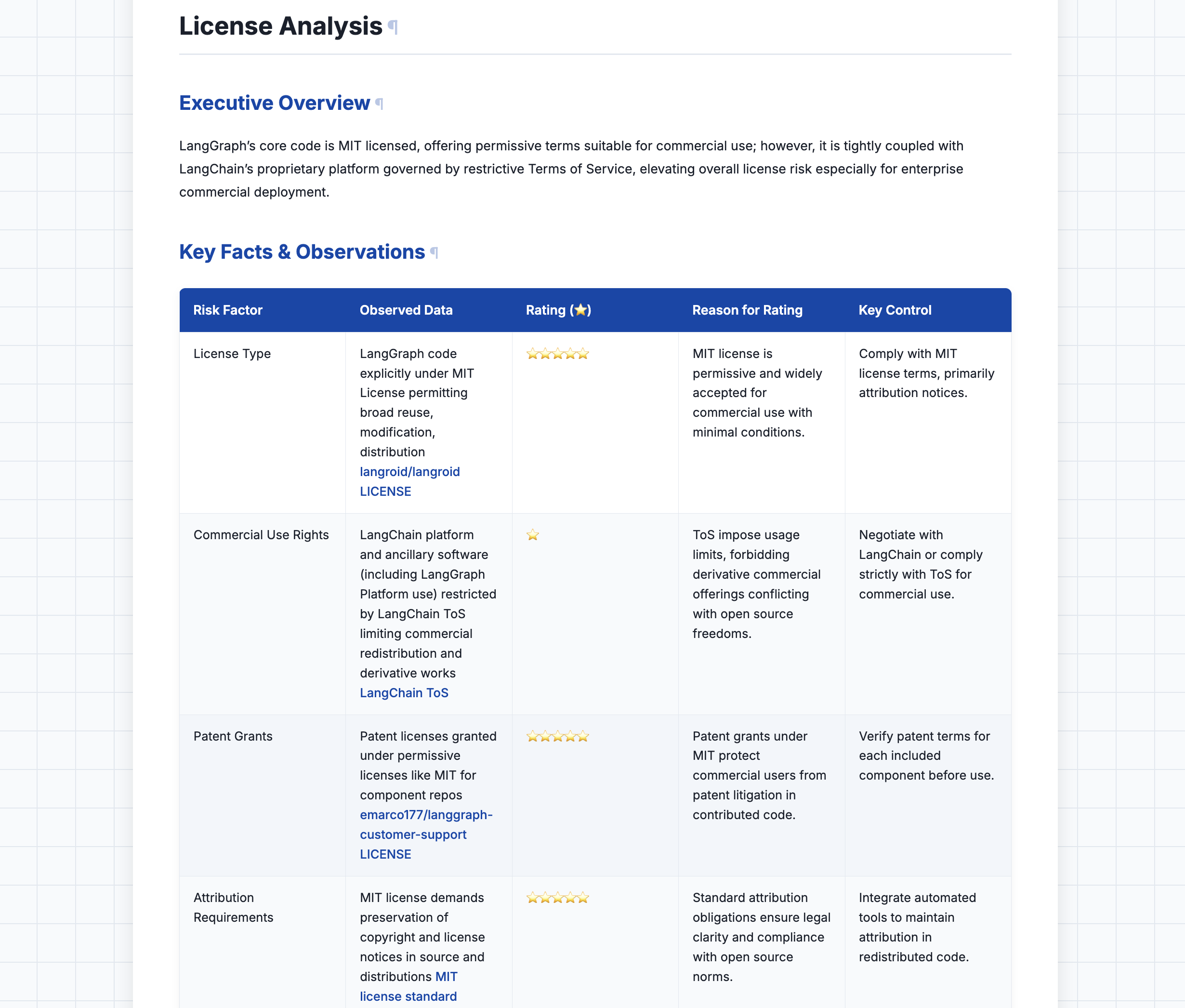}
    \caption{LangGraph License Analysis from the Generated Report, highlighting potential complexities arising from related Terms of Use.}
    \label{fig:langgraph-license-case}
\end{figure}

Our system identified that while LangGraph specifies an MIT license in its repository, a more comprehensive analysis revealed connections to LangChain\'s Terms of Use that potentially affect its licensing status. By understanding semantic relationships between documentation sources and interpreting licensing implications, the system provided a more holistic assessment than tools like the OpenSSF Scorecard, which primarily consider repository-level licensing information (see Figure~\ref{fig:langgraph-license-case}).

\subsubsection*{Regulatory Considerations: Browser Use}
\begin{figure}[h]
    \centering
    \includegraphics[width=\linewidth]{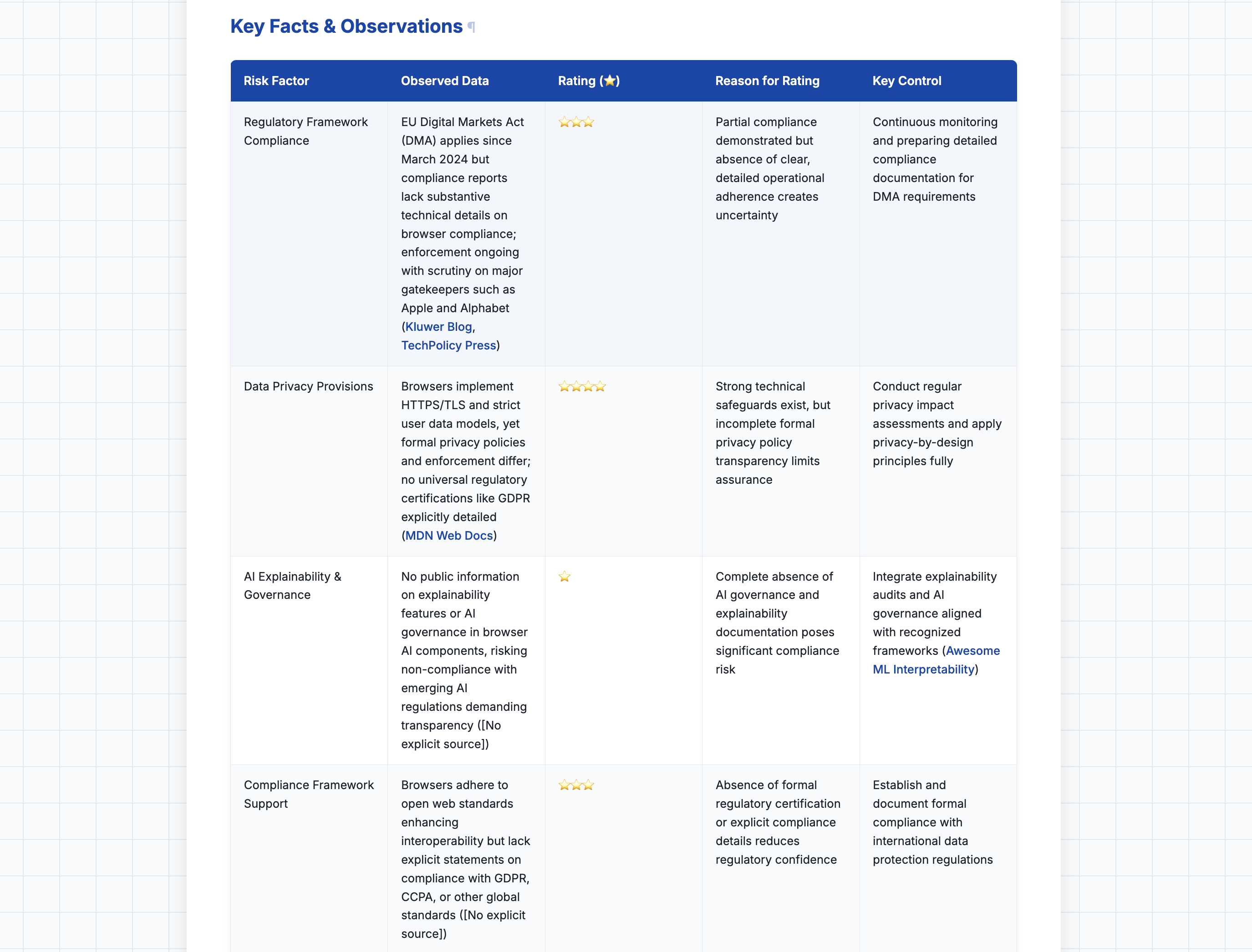}
    \caption{Browser Use Regulatory Analysis from the Generated Report, connecting library features to EU AI Act considerations.}
    \label{fig:browser-reg-case}
\end{figure}

For the Browser Use library, designed for web interaction tasks, \textsc{LibVulnWatch} linked its characteristics to emerging requirements under the EU AI Act. The system\'s ability to connect library functionality with regulatory frameworks enabled it to identify needs for clear documentation regarding data handling, agent capabilities, and potential risks, which are critical for compliance with high-risk AI system regulations (summarized in Figure~\ref{fig:browser-reg-case}). This showcases the value of language understanding in assessing alignment with evolving regulatory landscapes.

\subsubsection*{Security Analysis: JAX}
\begin{figure}[h]
    \centering
    \includegraphics[width=\linewidth]{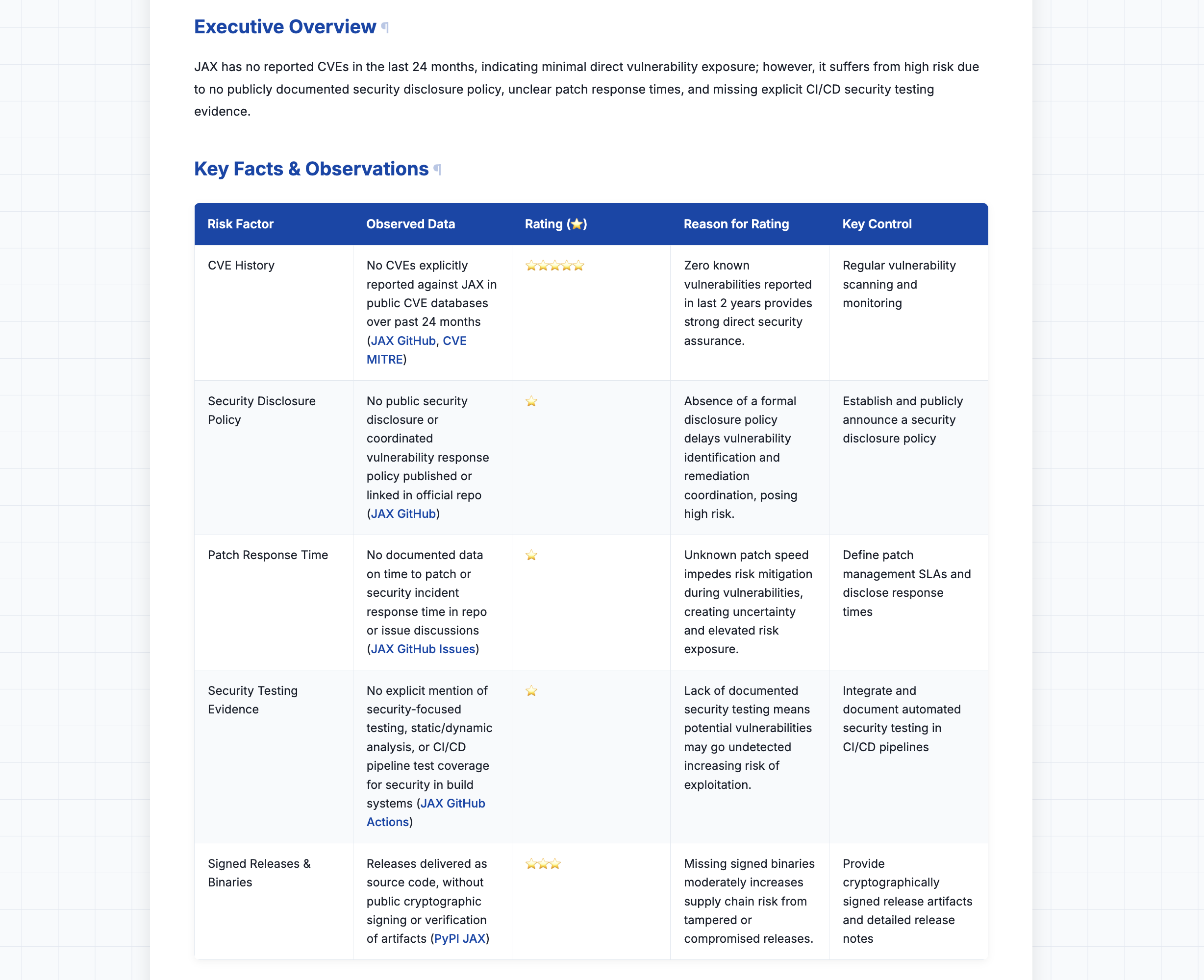}
    \caption{JAX Security Analysis from the Generated Report, noting absence of CVEs but also lack of explicit security CI.}
    \label{fig:jax-security-case}
\end{figure}

In the domain of security, \textsc{LibVulnWatch} correctly identified that the JAX library had no reported CVEs for the past two years. More importantly, through semantic analysis of GitHub Action links and repository structure, the system highlighted that JAX lacks an explicit, dedicated security Continuous Integration (CI) workflow, a subtle but important finding for long-term security posture that requires reasoning beyond simple pattern matching (Figure~\ref{fig:jax-security-case}).

\subsubsection*{Maintenance Analysis: vLLM}
\begin{figure}[h]
    \centering
    \includegraphics[width=\linewidth]{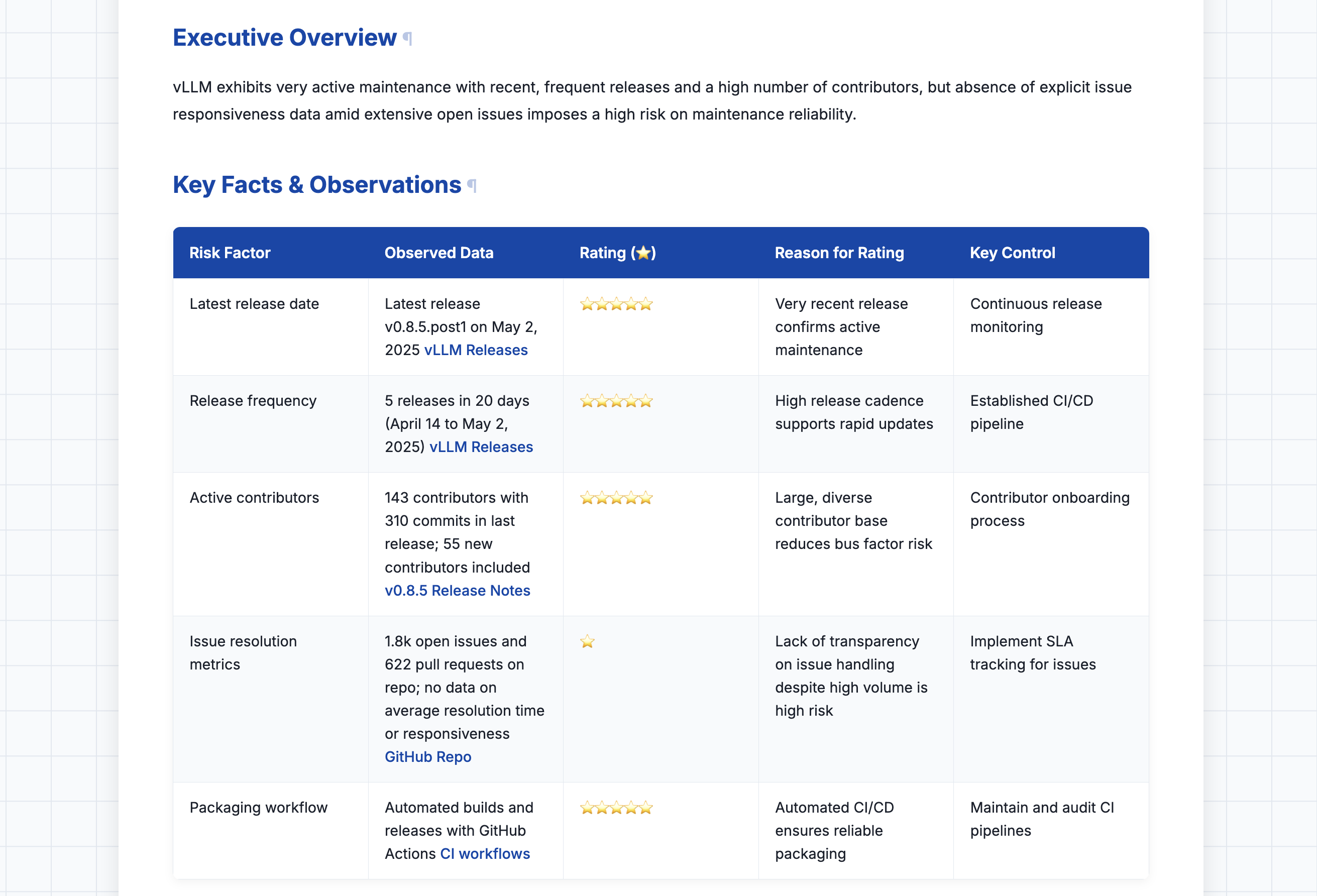}
    \caption{vLLM Maintenance Analysis from the Generated Report, summarizing repository activity trends.}
    \label{fig:vllm-main-case}
\end{figure}

For vLLM, an LLM inference and serving library, the system analyzed recent GitHub contributions, issue resolution times, and release frequency to assess its maintenance trends. By extracting and synthesizing temporal patterns from repository metadata, the system provided a quantitative overview of project activity, as shown in Figure~\ref{fig:vllm-main-case}, demonstrating how language models can integrate structured data analysis with contextual understanding.

\subsubsection*{Dependency Management: Huggingface Transformers}
\begin{figure}[h]
    \centering
    \includegraphics[width=\linewidth]{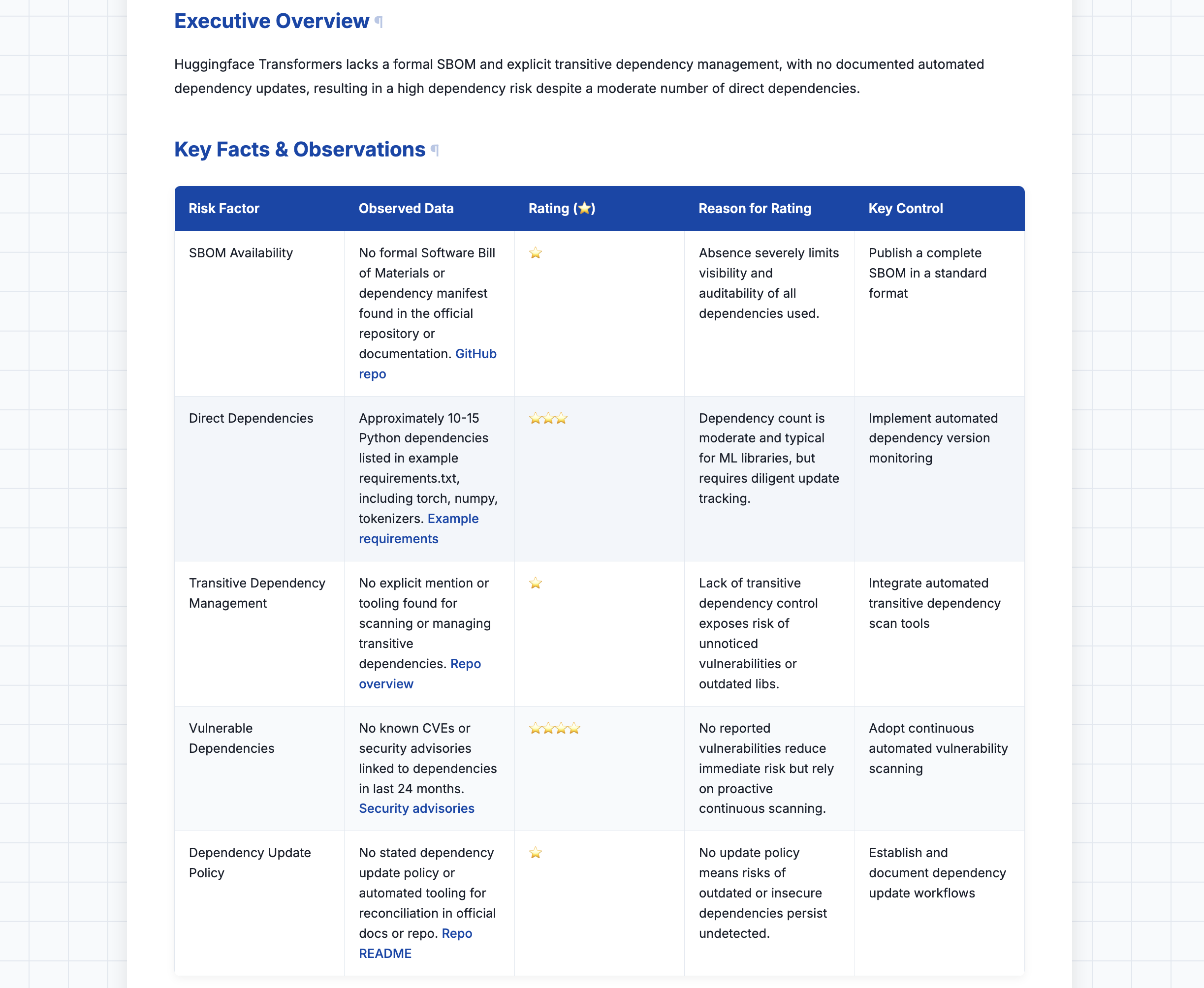}
    \caption{Huggingface Transformers Dependencies Analysis from the Generated Report.}
    \label{fig:hf-dep-case}
\end{figure}

\textsc{LibVulnWatch} examined the Huggingface Transformers library\'s dependency management practices. Leveraging its ability to interpret diverse information sources, the system evaluated the availability of a Software Bill of Materials (SBOM), analyzed stated policies regarding dependency updates, and assessed the overall approach to managing a complex dependency network. Figure~\ref{fig:hf-dep-case} illustrates a segment of this analysis, demonstrating how language-driven assessment can bridge technical details with governance requirements.

\section{Discussion and Future Work}
\label{sec:discussion}

Our findings reveal a critical gap: many technically advanced AI libraries exhibit significant shortcomings in enterprise readiness, particularly in supply chain security and regulatory preparedness (Section~\ref{subsec:domain-findings-patterns}). This underscores a pressing need for more nuanced assessment methodologies. The agent-based approach we introduced (Section~\ref{subsec:agentic-graph}), rooted in language understanding, proved effective in identifying complex vulnerabilities—such as RCEs, supply chain flaws, and governance gaps—that elude conventional checks. The substantial Novelty Yield achieved (Table~\ref{tab:coverage-summary}, Section~\ref{subsec:benchmark-analysis}) quantifies this unique contribution, demonstrating how NLP can uncover critical risks requiring deep contextual interpretation, a finding further supported by the patterns detailed in Section~\ref{subsec:domain-findings-patterns}.

Benchmarking our system (Section~\ref{sec:results}) against established tools like the OpenSSF Scorecard provides a crucial perspective. While the observed Baseline Alignment (Section~\ref{subsec:benchmark-analysis}, Table~\ref{tab:coverage-summary}) confirms our method's capacity to recognize standard risk indicators, the consistent generation of novel insights highlights the added value of recontextualizing NLP for specialized domains. The variations in alignment and novelty across library categories (Table~\ref{tab:risk-assessment-scores-main}, Section~\ref{subsec:evaluated-libraries}) suggest that a library's functional niche and maturity, rather than mere complexity, influence its risk profile when assessed through this deeper, language-aware lens.

This work offers a clear demonstration of how advanced language understanding capabilities can transform risk assessment methodologies, moving beyond traditional rule-based paradigms (Section~\ref{sec:methodology}). The system's proficiency in interpreting diverse documentation, synthesizing disparate information, and reasoning about nuanced implications (Figure~\ref{fig:assessment-architecture}) facilitates a depth of analysis previously unattainable with conventional tools. Crucially, this approach enables the identification of emergent, cross-cutting patterns, such as systemic deficiencies in regulatory alignment (Section~\ref{subsec:domain-findings-patterns}), thereby offering insights into broader ecosystemic challenges that demand interdisciplinary attention.

Looking ahead, our research points towards several avenues for intensifying NLP's impact in this and related domains. Enhancing the semantic interpretation of code and API interactions, grounded in our current risk framework (Section~\ref{subsec:risk-framework}), promises more precise intra-implementation vulnerability detection. The successful application of this NLP-driven framework (Section~\ref{sec:methodology}) to software assessment strongly motivates its adaptation to other complex ecosystems, such as healthcare informatics or financial technologies, where similar governance and risk assessment challenges persist. Further exploration of few-shot adaptation could democratize such deep assessment capabilities. Ultimately, integrating structured verification techniques with the contextual reasoning inherent in language models could address current limitations while amplifying the discovery of impactful, novel risks, as evidenced by our Novelty Yield results (Table~\ref{tab:coverage-summary}, Section~\ref{subsec:benchmark-analysis}).

Collectively, these contributions signal a paradigm shift: viewing the evaluation of complex systems not merely as a static analysis task, but as a dynamic knowledge synthesis challenge. This perspective directly leverages recent breakthroughs in language understanding and structured reasoning. By effectively bridging NLP with the distinct domain of software governance, \textsc{LibVulnWatch} (Section~\ref{sec:methodology}, Section~\ref{sec:results}) provides not only actionable insights for AI library evaluation but also a robust, transferable methodology for tackling multifaceted governance and risk assessment problems across diverse disciplinary boundaries.
\section{Limitations}
\label{sec:limitations}

Despite the capabilities of \textsc{LibVulnWatch}, several limitations warrant discussion, offering avenues for future research and refinement.

\paragraph{Refined Agent Capabilities and Scope} While \textsc{LibVulnWatch} demonstrates broad alignment with the OpenSSF Scorecard (as discussed in Section~\ref{subsec:benchmark-analysis}), its agentic reasoning did not consistently capture all specific checklist items, such as the presence of binary artifacts or formal contributor agreements. This suggests that for comprehensive coverage of all standard security hygiene factors, future iterations could benefit from incorporating more specialized, non-agentic tools or targeted heuristics for these highly structured data points, complementing the agent\'s deep analysis of more nuanced risks.

\paragraph{Dynamic Nature of Open-Source and Information Availability} The accuracy and completeness of \textsc{LibVulnWatch} assessments are intrinsically tied to the availability and quality of public information concerning the target libraries. As open-source projects evolve rapidly, any assessment inherently represents a snapshot in time (e.g., data for this paper reflects May 2025, a point also noted in Section~\ref{sec:discussion}). While continuous monitoring via the planned public leaderboard (Section~\ref{subsec:agentic-graph}) aims to mitigate the staleness of information, the depth of analysis will always be constrained by what projects choose to disclose publicly and the recency of indexed information by search APIs.

\paragraph{LLM Dependence and Evaluation Robustness} \textsc{LibVulnWatch} leverages the capabilities of LLMs (specifically \texttt{gpt-4.1-mini}) for complex information extraction and synthesis. Consequently, the quality and consistency of assessments can be influenced by the LLM\'s inherent knowledge envelope, reasoning limitations, potential training data biases, and sensitivity to prompt engineering, as acknowledged in Section~\ref{sec:discussion}. Although our framework emphasizes evidence-backed findings and structured reporting to mitigate subjectivity and ensure verifiability (Section~\ref{subsec:agentic-graph}), future work could explore ensembles of diverse LLMs, more rigorous calibration of prompt variance, or techniques for explicitly surfacing LLM uncertainty in assessments.

\paragraph{Scalability and Resource Implications for Deep, Continuous Analysis} Performing deep, source-grounded analysis for a large number of libraries on a continuous basis presents computational resource considerations. While individual library assessments with \texttt{gpt-4.1-mini} are relatively cost-effective (approx. \$0.10 per library, as detailed in Section~\ref{subsec:agentic-graph}), scaling this to thousands of libraries with high frequency would necessitate significant infrastructure. Future optimizations might involve adaptive assessment depths based on library criticality or observed change frequency, or the development of more efficient caching mechanisms for retrieved evidence.

\paragraph{Ecosystem-Level Constraints on Assessment Depth} A significant constraint, external to \textsc{LibVulnWatch} itself, is the current state of documentation within the open-source AI ecosystem. The pervasive lack of comprehensive and standardized documentation regarding regulatory compliance (e.g., GDPR, AI Act alignment), detailed privacy practices, and robust model/data explainability inherently limits the depth and certainty of assessments in these critical governance domains. While our system is designed to identify such gaps (a pattern noted in Section~\ref{subsec:domain-findings-patterns})—which itself is a valuable finding—it cannot create information that does not exist. This limitation underscores a broader need for community-driven standards and improved transparency from library developers to enable more thorough governance evaluations.
\section{Ethical Considerations}
\label{sec:ethical_considerations}

The development and deployment of \textsc{LibVulnWatch} raise several ethical considerations that we have aimed to address throughout its design and proposed usage.

\paragraph{Responsible Disclosure and Vulnerability Reporting} As stated in our methodology (Section~\ref{subsec:agentic-graph}), \textsc{LibVulnWatch} is designed to identify potential vulnerabilities in open-source AI libraries. We are committed to responsible disclosure practices. For any new, previously non-public vulnerabilities, particularly critical ones such as the RCEs mentioned in our results (Section~\ref{subsec:benchmark-analysis}), our protocol involves adhering to the ACL Co-ordinated Disclosure Policy. This includes contacting the developers of the affected library privately, providing them with the necessary details, and allowing a minimum 30-day period for them to address the issue before any public disclosure of the specific, novel vulnerability details. All such communications and their timelines would be documented herein or in a publicly available appendix upon final publication if such instances arise during ongoing or future assessments.

\paragraph{Potential for Misuse} While \textsc{LibVulnWatch} aims to improve the security and governance of the AI ecosystem by highlighting risks, any tool that identifies vulnerabilities could potentially be misused by malicious actors. To mitigate this, our public leaderboard (as referenced in Section~\ref{subsec:agentic-graph}) focuses on aggregated, governance-aligned scores and known risk patterns rather than detailing zero-day exploits. The primary goal is to incentivize proactive security improvements and inform developers and users, with responsible disclosure handling specific sensitive findings. Furthermore, the types of vulnerabilities it highlights (e.g., missing SBOMs, licensing issues, gaps in regulatory documentation) are often systemic issues that benefit from public awareness to drive broader improvements.

\paragraph{LLM Capabilities, Biases, and Reproducibility} The assessment quality of \textsc{LibVulnWatch} is inherently linked to the capabilities and potential biases of the underlying Large Language Model (LLM), \texttt{gpt-4.1-mini}, as noted in our limitations (Section~\ref{sec:discussion}). While we employ engineered prompts and a structured, evidence-based framework (Sections~\ref{subsec:risk-framework} and~\ref{subsec:agentic-graph}) to guide the LLM and ensure verifiability (e.g., quantification mandate, evidence requirement), the interpretation and synthesis performed by the LLM may still be subject to its training data biases or inherent limitations. We strive for transparency by detailing our methodology, including the use of specific LLM agents and prompts (though full prompt details are beyond the scope of this paper, the principles are outlined). The generated reports, with direct citations to evidence, are designed to be reproducible and allow for independent verification of findings.

\paragraph{Data Privacy} \textsc{LibVulnWatch} is designed to assess publicly available open-source AI libraries. The data sources it utilizes, as described in Section~\ref{subsec:agentic-graph}, include public code repositories, official documentation, security databases, and information retrieved via public web search APIs. The system does not require access to private codebases or non-public user data, minimizing direct data privacy risks related to proprietary information.

\paragraph{Impact of Public Ranking and Scoring} Publishing a leaderboard with risk scores for AI libraries can have a significant societal impact. Our intention is to foster transparency, accountability, and drive improvements in the security and governance of the AI software supply chain. However, we recognize that scores could be misinterpreted or place undue pressure on developers of libraries that score lower. To mitigate this, \textsc{LibVulnWatch} emphasizes a multi-dimensional assessment across five domains (Section~\ref{subsec:risk-framework}), detailed justifications for scores, and evidence-backed findings, rather than a single opaque metric. The OpenSSF Scorecard benchmarking (Section~\ref{subsec:baseline-metrics}) also provides a recognized baseline for comparison. We believe the benefits of increased transparency and informed decision-making for users and developers outweigh the potential downsides, especially given the critical nature of these libraries in AI systems.

\paragraph{Fairness and Objectivity} We have designed the assessment framework to be as objective as possible by mandating structured reporting, quantification of metrics, and direct evidence for all claims (Section~\ref{subsec:agentic-graph}). The risk rating criteria (Section~\ref{subsec:quantification}) are predefined to ensure consistency across evaluations. While the LLM introduces a layer of interpretation, the requirement for verifiable evidence aims to ground the assessments in factual data.

We believe that by adhering to these principles, \textsc{LibVulnWatch} can serve as a valuable and ethical tool for enhancing the trustworthiness of the open-source AI ecosystem. 

\bibliography{references}

\onecolumn
\newpage
\appendix
\section{Appendix}
\label{appendix}

\subsection{Interactive Leaderboard Interface and Implementation}
\label{subsec:LeaderboardInterface}
\begin{figure}[htbp]
    \centering
    \begin{minipage}[b]{0.48\textwidth}
        \centering
        \includegraphics[width=\textwidth]{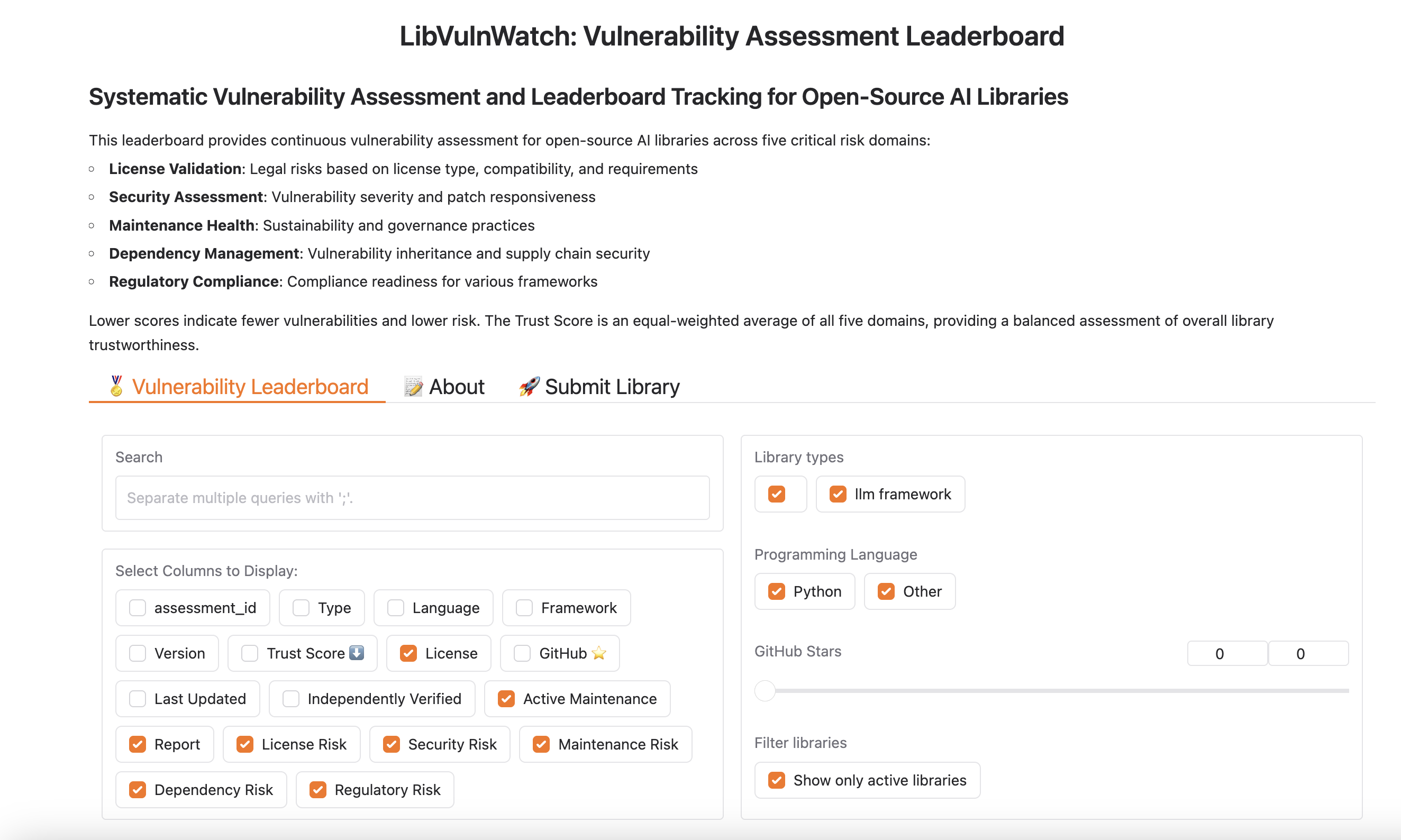}
        \caption{The main \textsc{LibVulnWatch} leaderboard view, showing search and filtering options for assessed AI libraries across five risk domains.}
        \label{app:leaderboard-main}
    \end{minipage}
    \hfill
    \begin{minipage}[b]{0.48\textwidth}
        \centering
        \includegraphics[width=\textwidth]{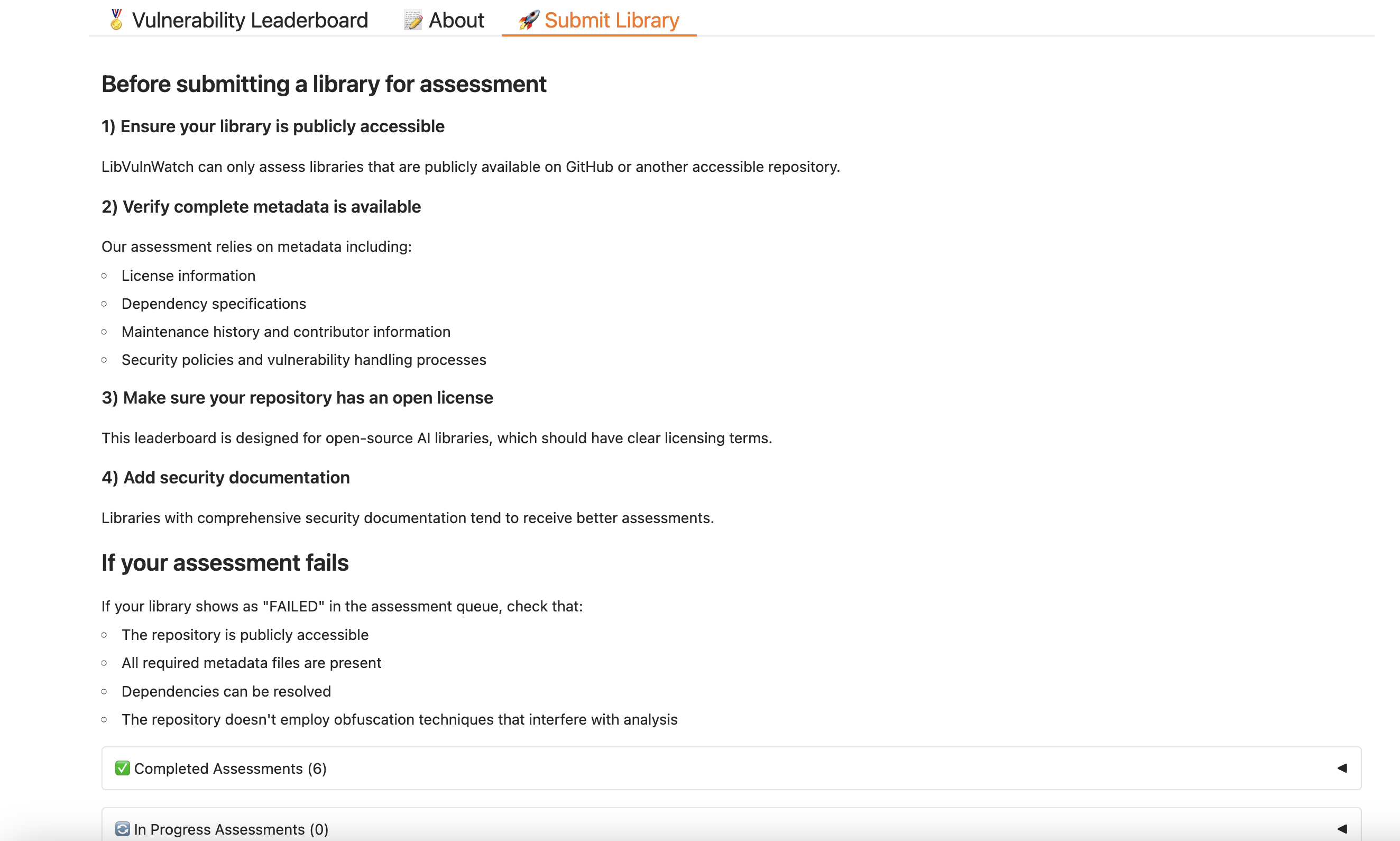}
        \caption{Guidelines and prerequisites for submitting a new library for assessment on the \textsc{LibVulnWatch} platform.}
        \label{app:leaderboard-submit-guidelines}
    \end{minipage}
    
    \vspace{2 em} 
    
    \begin{minipage}[b]{0.48\textwidth}
        \centering
        \includegraphics[width=\textwidth]{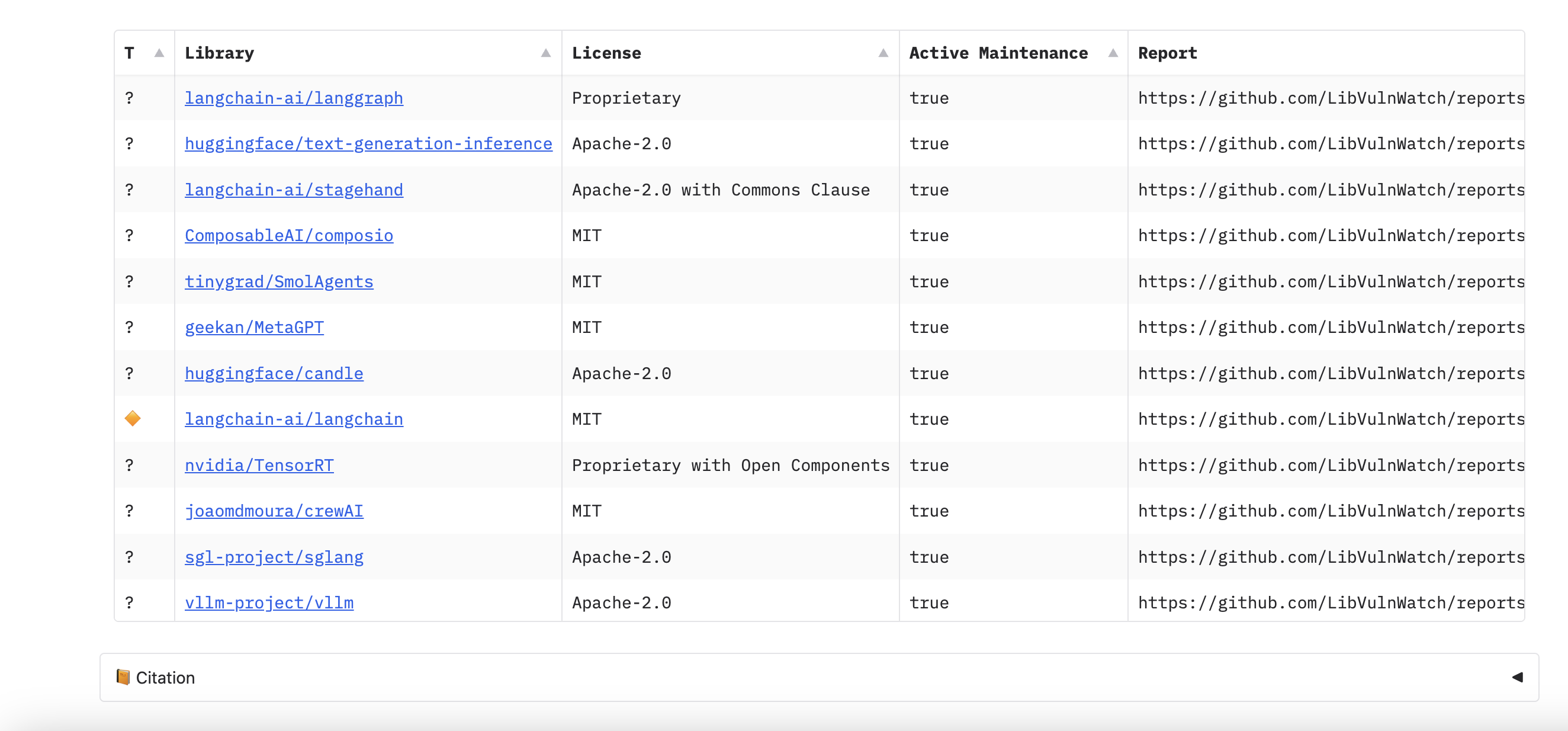}
        \caption{Tabular display of assessed libraries, including details such as license, maintenance status, and direct links to individual vulnerability reports.}
        \label{app:leaderboard-table}
    \end{minipage}
    \hfill
    \begin{minipage}[b]{0.48\textwidth}
        \centering
        \includegraphics[width=\textwidth]{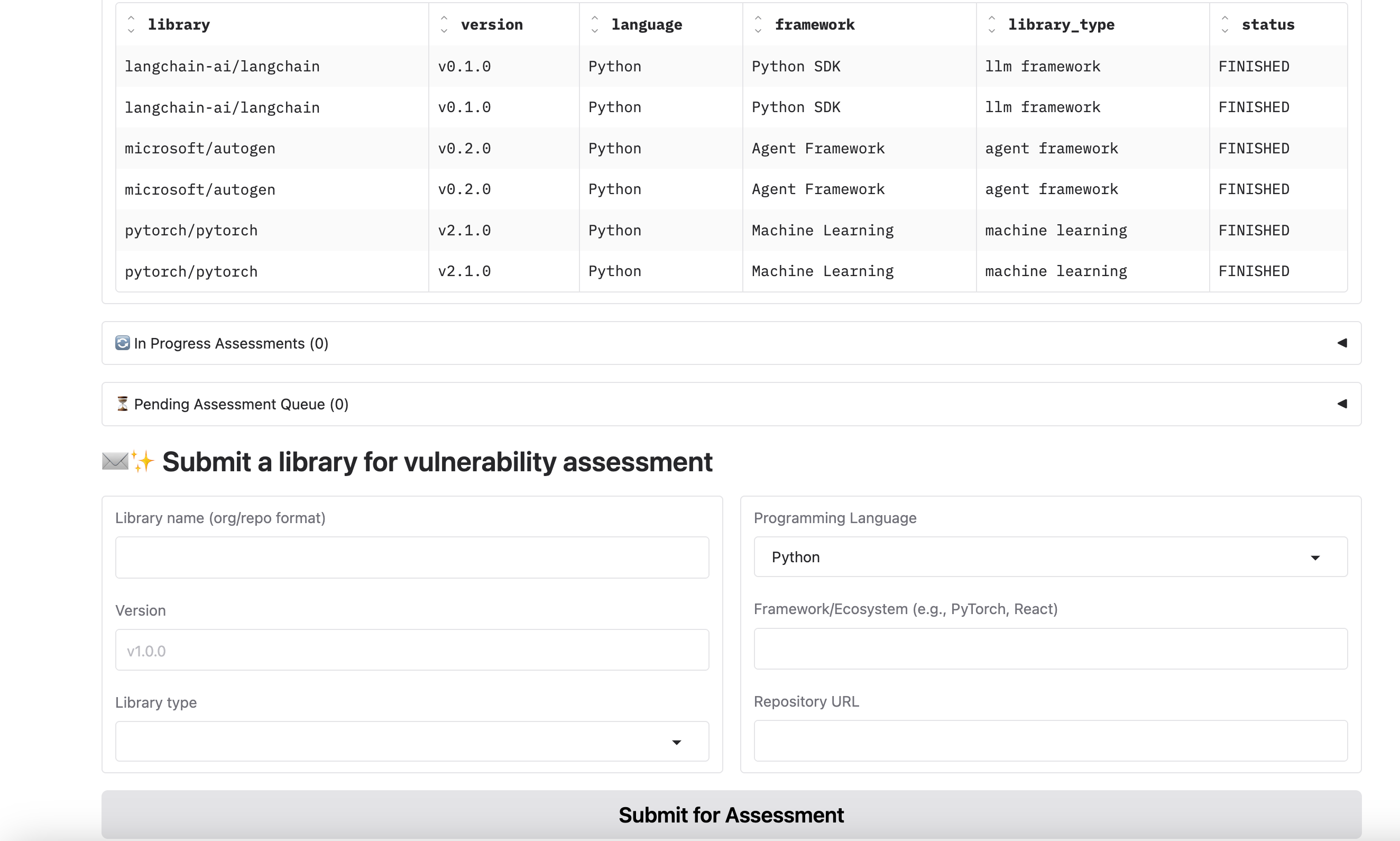}
        \caption{The \textsc{LibVulnWatch} interface for submitting a new open-source AI library for vulnerability assessment and inclusion in the leaderboard.}
        \label{app:leaderboard-submit-form}
    \end{minipage}
\end{figure}

This subsection describes the \textsc{LibVulnWatch} vulnerability assessment leaderboard and presents screenshots of its key functionalities. The leaderboard is implemented as an interactive web application using Gradio~ and is publicly deployed on Hugging Face Spaces. It allows users to search, filter, and view detailed vulnerability assessment reports for a wide range of open-source AI libraries. Users can also find guidelines and submit new libraries for assessment through this interface. The figures below illustrate the main views of the leaderboard, including search and filtering capabilities (Figure~\ref{app:leaderboard-main}), library submission guidelines (Figure~\ref{app:leaderboard-submit-guidelines}), the tabular display of assessed libraries with links to reports (Figure~\ref{app:leaderboard-table}), and the new library submission form (Figure~\ref{app:leaderboard-submit-form}).

\subsection{Agent Prompts}
\label{subsec:AgentPrompts}
This section details the core instruction sets (prompts) provided to the various specialized agents within the \textsc{LibVulnWatch} system. These prompts guide the agents in their respective tasks of planning, querying, writing, and evaluating risk assessment information.

\subsubsection{Initial Query Formulation for Report Planning}
\begin{lstlisting}[caption={Initial Query Formulation for Report Planning. This agent generates initial search queries to gather context for planning the overall report structure.}, basicstyle=\ttfamily\tiny, breaklines=true, postbreak=\mbox{{\color{red}$\hookrightarrow$}\space}, inputencoding=utf8, literate={🚨}{{[ALERT]}}{7} {⭐}{{$\star$}}{1}]
You are performing comprehensive open source risk management assessment following industry best practices.

<Library input>
{topic}
</Library input>

<Report organization>
{report_organization}
</Report organization>

<Task>
Your goal is to generate {number_of_queries} web search queries that will gather comprehensive information for assessing the risks of this open source library according to enterprise security standards.

IMPORTANT: The library input may be either a library name (e.g., "TensorFlow", "React") or a repository URL (e.g., "https://github.com/tensorflow/tensorflow"). Adjust your queries accordingly.

<High-Quality Source Guidelines>
Prioritize authoritative and reliable sources by targeting queries toward:
- Official documentation (GitHub repos, project websites, official guides)
- Security databases (NVD, CVE records, security bulletins)
- Industry research (research papers, security firm reports)
- Regulatory bodies (NIST, ISO, CIS documentation)
- Technical forums with verification (StackOverflow with high votes)

Avoid low-quality sources like:
- General blogs without technical expertise
- Marketing materials
- Outdated repositories (>2 years without updates)
- Non-technical news articles

Use site: operators to target specific high-quality domains (e.g., site:github.com, site:nvd.nist.gov).
</High-Quality Source Guidelines>

The queries should comprehensively cover these key risk areas:

1. LICENSE VALIDATION:
   - License type (MIT, Apache 2.0, GPL, etc.)
   - Commercial use compatibility
   - License history and changes
   - Attribution requirements
   - Patent grant provisions

2. SECURITY ASSESSMENT:
   - Common Vulnerabilities and Exposures (CVEs)
   - Security patch frequency and responsiveness
   - Vulnerability scanning reports
   - OWASP dependency risks
   - Historical security incidents

3. MAINTENANCE HEALTH:
   - Release frequency and consistency
   - Number of active contributors (current vs. historical)
   - Issue response time metrics
   - Pull request acceptance rate
   - Governance model (individual, community, foundation)

4. DEPENDENCY MANAGEMENT:
   - Software Bill of Materials (SBOM) availability
   - Transitive dependency tracking
   - Dependency update policies
   - Supply chain security measures
   - CI/CD integration for dependency scanning

5. REGULATORY COMPLIANCE:
   - Explainability requirements (especially for AI libraries)
   - Industry-specific regulatory frameworks applicable
   - Data privacy implications
   - Export control restrictions
   - Audit readiness documentation

Make the queries specific, technical, and designed to retrieve quantifiable metrics from authoritative sources wherever possible. Use site: operators to target specific high-quality domains when appropriate.
</Task>

<Format>
Call the Queries tool
</Format>
\end{lstlisting}

\subsubsection{Report Structure Planning Instructions}
\begin{lstlisting}[caption={Report Structure Planning Instructions. This agent generates the structured plan for the report, outlining the sections to be created.}, basicstyle=\ttfamily\tiny, breaklines=true, postbreak=\mbox{{\color{red}$\hookrightarrow$}\space}, inputencoding=utf8, literate={🚨}{{[ALERT]}}{7} {⭐}{{$\star$}}{1}]
I want a comprehensive open source risk assessment report that meets enterprise governance standards and regulatory compliance requirements.

<Library input>
The library to assess is:
{topic}
</Library input>

<Report organization>
The report should follow this organization:
{report_organization}
</Report organization>

<Context>
Here is context to use to plan the sections of the risk assessment report:
{context}
</Context>

<Task>
Generate a detailed structure for an enterprise-grade open source risk assessment report on the provided library.

IMPORTANT: The library input may be either a library name (e.g., "TensorFlow", "React") or a repository URL (e.g., "https://github.com/tensorflow/tensorflow"). Identify the specific library from the input.

Your plan should include specialized sections that cover ALL of the following risk domains based on industry best practices:

1. KEY RISK DOMAINS (each requiring full assessment as separate sections):
   - LICENSE ANALYSIS - Terms, compatibility, patent provisions
   - SECURITY ASSESSMENT - CVE history, patch frequency, testing
   - MAINTENANCE INDICATORS - Release cadence, contributors, support
   - DEPENDENCY MANAGEMENT - SBOM, transitive risks, updates
   - REGULATORY CONSIDERATIONS - Compliance frameworks, explainability

NOTE:
- The EXECUTIVE SUMMARY will be generated automatically after all sections are written, so DO NOT include it in your section list.
- RISK MITIGATION RECOMMENDATIONS will be included in the Executive Summary, so DO NOT create it as a separate section.

Each section should have the fields:
- Name - Name for this section of the report.
- Description - Brief overview of what this section assesses.
- Research - Whether to perform web research for this section. IMPORTANT: All main sections MUST have Research=True.
- Content - The content of the section, which you will leave blank for now.

Ensure the structure focuses on quantifiable metrics and evidence-based assessment rather than general descriptions. The report should be highly actionable, non-redundant, and concise.
</Task>

<Feedback>
Here is feedback on the report structure from review (if any):
{feedback}
</Feedback>

<Format>
Call the Sections tool
</Format>
\end{lstlisting}

\subsubsection{Domain-Specific Query Formulation Instructions}
\begin{lstlisting}[caption={Domain-Specific Query Formulation Instructions. This agent generates specific search queries for a given section of the report.}, basicstyle=\ttfamily\tiny, breaklines=true, postbreak=\mbox{{\color{red}$\hookrightarrow$}\space}, inputencoding=utf8, literate={🚨}{{[ALERT]}}{7} {⭐}{{$\star$}}{1}]
You are an enterprise security analyst specializing in open source risk governance and compliance.

<Library input>
{topic}
</Library input>

<Section topic>
{section_topic}
</Section topic>

<Task>
Generate {number_of_queries} highly specific search queries to gather comprehensive data for assessing the open source risks of this library, focusing specifically on {section_topic}.

IMPORTANT: The library input may be either a library name (e.g., "TensorFlow", "React") or a repository URL (e.g., "https://github.com/tensorflow/tensorflow"). Always include the library name explicitly in your queries.

<Advanced GitHub Data Extraction>
Since we do not have API access, use these specialized search patterns to extract public repository metrics:

1. For contributor metrics:
   - "[Library] github.com/[org]/[repo]/graphs/contributors" (finds contributor pages)
   - "[Library] [org]/[repo] number of contributors [year]" (finds specific counts)
   - "[Library] [org]/[repo] top contributors" (finds lead maintainer information)

2. For issue statistics:
   - "[Library] github.com/[org]/[repo]/issues?q=is:issue+is:open+sort:updated-desc" (finds open issues)
   - "[Library] github.com/[org]/[repo]/issues?q=is:issue+is:closed" (finds closed issues)
   - "[Library] average issue resolution time" (finds resolution metrics)

3. For release history:
   - "[Library] github.com/[org]/[repo]/releases" (finds release pages)
   - "[Library] latest release version number date" (finds current version)
   - "[Library] release frequency [year]" (finds release cadence)

4. For security practices:
   - "[Library] github.com/[org]/[repo]/security/advisories" (finds security advisories)
   - "[Library] github.com/[org]/[repo]/blob/master/SECURITY.md" (finds security policies)
   - "[Library] CVE [year] vulnerability" (finds published vulnerabilities)

5. For dependency information:
   - "[Library] github.com/[org]/[repo]/blob/master/requirements.txt" (finds Python dependencies)
   - "[Library] github.com/[org]/[repo]/blob/master/package.json" (finds JS dependencies)
   - "[Library] github.com/[org]/[repo]/network/dependencies" (finds dependency graphs)

6. For license details:
   - "[Library] github.com/[org]/[repo]/blob/master/LICENSE" (finds license file)
   - "[Library] github.com/[org]/[repo]/blob/master/LICENSE.md" (alternative license file)
   - "[Library] license type changed history" (finds license changes)
</Advanced GitHub Data Extraction>

<High-Quality Source Guidelines>
Prioritize authoritative and reliable sources by targeting queries toward:
- Official documentation (GitHub repos, project websites, official guides)
- Security databases (NVD, CVE records, security bulletins)
- Industry research (research papers, security firm reports)
- Regulatory bodies (NIST, ISO, CIS documentation)
- Technical forums with verification (StackOverflow with high votes)

Avoid low-quality sources like:
- General blogs without technical expertise
- Marketing materials
- Outdated repositories (>2 years without updates)
- Non-technical news articles

Your queries should specifically target these high-quality sources when possible.
</High-Quality Source Guidelines>

Based on the section topic, craft specialized queries from these categories:

LICENSE ANALYSIS:
- "[Library] license type commercial use compatibility site:github.com OR site:opensource.org"
- "[Library] license change history site:github.com/[org]/[repo]"
- "[Library] patent grant provisions license text"
- "[Library] license compliance requirements site:spdx.org OR site:github.com"
- "[Library] GPL/LGPL/AGPL compatibility analysis"
- "[Library] attribution requirements license text site:opensource.org"

SECURITY ASSESSMENT:
- "[Library] CVE history last 3 years site:nvd.nist.gov OR site:cve.mitre.org"
- "[Library] security vulnerabilities mitigated site:github.com/[org]/[repo]/security"
- "[Library] CVSS score recent vulnerabilities site:nvd.nist.gov"
- "[Library] security disclosure policy site:github.com/[org]/[repo]"
- "[Library] security patch response time average"
- "[Library] supply chain security scorecard"

MAINTENANCE HEALTH:
- "[Library] release frequency metrics site:github.com/[org]/[repo]/releases"
- "[Library] active contributors count trend site:github.com/[org]/[repo]/graphs/contributors"
- "[Library] issue resolution time average site:github.com/[org]/[repo]/issues"
- "[Library] pull request acceptance rate site:github.com/[org]/[repo]/pulls"
- "[Library] documentation quality assessment site:github.com/[org]/[repo]/wiki"
- "[Library] governance foundation or company site:github.com OR site:[official-site]"

DEPENDENCY MANAGEMENT:
- "[Library] SBOM availability CycloneDX or SPDX site:github.com/[org]/[repo]"
- "[Library] transitive dependencies count analysis"
- "[Library] dependency vulnerability scanning site:github.com/[org]/[repo]/security/dependabot"
- "[Library] dependency freshness policy site:github.com/[org]/[repo]"
- "[Library] CI/CD dependency scanning integration site:github.com/[org]/[repo]/.github/workflows"
- "[Library] vulnerable dependencies percentage report"

REGULATORY COMPLIANCE:
- "[Library] regulatory compliance frameworks site:[official-site] OR site:github.com/[org]/[repo]"
- "[Library] explainability for AI models documentation site:github.com/[org]/[repo]"
- "[Library] data privacy implications GDPR CCPA CPRA site:github.com/[org]/[repo]"
- "[Library] export control classification ECCN"
- "[Library] NIST SSDF compatibility assessment"
- "[Library] audit readiness documentation site:github.com/[org]/[repo]"

<Data Extraction Instructions>
For each query, focus on extracting specific numerical metrics:
- Always search for EXACT numbers when available: "X contributors" not "many contributors"
- Look for timestamps and dates: "Last release: March 15, 2024" not "recent release"
- Search for explicit vulnerability counts: "3 CVEs in 2023" not "some vulnerabilities"
- Seek percentages and ratios: "85% test coverage" not "good test coverage"

For repositories, use google dorks to find specific file content:
- Use 'inurl:github.com/[org]/[repo] filetype:md SECURITY' to find security documentation
- Use 'inurl:github.com/[org]/[repo] "license"' to find license information
- Use 'inurl:github.com/[org]/[repo] "requirements.txt" OR "package.json"' to find dependencies
</Data Extraction Instructions>

Generate queries that return quantitative metrics, statistical data, and factual evidence from authoritative sources. Use site: operators when appropriate to target specific high-quality domains.
</Task>

<Format>
Call the Queries tool
</Format>
\end{lstlisting}

\subsubsection{Draft Findings Generation Instructions for Report Sections}
\begin{lstlisting}[caption={Draft Findings Generation Instructions for Report Sections. This agent synthesizes information from web search results to write a specific section of the report, adhering to strict formatting and citation requirements.}, basicstyle=\ttfamily\tiny, breaklines=true, postbreak=\mbox{{\color{red}$\hookrightarrow$}\space}, inputencoding=utf8, literate={🚨}{{[ALERT]}}{7} {⭐}{{$\star$}}{1}]
Write a highly focused assessment of open source risk.

<Task>
1. Analyze the library based on the section name and topic.
2. Focus ONLY on observed facts with proper citations.
3. Use the most concise format possible while addressing all key risk factors.
4. IMPORTANT: For each risk factor, assign at least one HIGH risk rating if evidence justifies it. Never rate all factors as only Low/Medium.
</Task>

<Streamlined Structure>
## [Section Name]

### Executive Overview
[1 sentence summary of risk level and justification]

### 🚨 Emergency Issues
<span style="color:red">
**Critical Issue**: [Most serious high-risk finding with citation link for critical information only](url)
</span>

### Key Facts & Observations
| Risk Factor | Observed Data | Rating (⭐) | Reason for Rating | Key Control |
|-------------|---------------|------------|-------------------|-------------|
| [Factor 1]  | [Specific metric/fact with citation link](url) | ⭐⭐⭐⭐⭐ | [Why this is low risk] | [Solution]  |
| [Factor 2]  | [Specific metric/fact with citation link](url) | ⭐⭐⭐ | [Why this is medium risk] | [Solution]  |
| [Factor 3]  | [Specific metric/fact with citation link](url) | ⭐ | [Why this is high risk] | [Solution]  |
</Streamlined Structure>

<Coverage Requirements>
Based on your section topic, address ALL relevant key concepts:

LICENSE ANALYSIS:
- License type (MIT, Apache, GPL, etc.) with version
- Commercial use & distribution rights
- Patent grant provisions
- Attribution requirements
- Conformance with open source compliance standards

SECURITY ASSESSMENT:
- CVEs in past 24 months (count, severity)
- Security disclosure policy existence
- Response time for security issues
- Security testing evidence (CI/CD test coverage)
- Released binaries or signed artifacts and release notes

MAINTENANCE INDICATORS:
- Latest release date
- Release frequency (releases per month/year)
- Active contributor count (diversity and organizational backing)
- Issue resolution metrics (recent commit activity and issue engagement details)
- Packaging workflow for publishing

DEPENDENCY MANAGEMENT:
- SBOM availability (Yes/No, format)
- Direct dependency count
- Transitive dependency management
- Vulnerable dependency count
- Existence of dependency update tools/policies

REGULATORY CONSIDERATIONS:
- Compliance frameworks supported
- Explainability features for AI/ML
- Data privacy provisions
- Audit documentation availability
- AI governance and key AI regulations

CRITICAL: Ensure EVERY metric has a specific value, NOT general statements.
</Coverage Requirements>

<Writing Guidelines>
- Extract the library name from the input (may be name or repository URL)
- Use ONLY observed facts and metrics with citations:
  - "Last release: March 15, 2024" not "recent release"
  - "243 active contributors" not "many contributors"
  - "No CVEs in past 24 months" not "good security record"

- STRICT CITATION REQUIREMENTS:
  - ONLY make claims that are EXPLICITLY stated in the source material
  - DO NOT infer, assume, or extrapolate beyond what's directly stated in the sources
  - If source material does not explicitly mention a metric, acknowledge this as "No data available on X" and rate accordingly
  - Maintain clear traceability between each claim and the exact source
  - For missing but important information, indicate "Not specified in documentation" rather than guessing

- CITATION FORMAT AND FREQUENCY:
  - ONLY use inline markdown hyperlinks for direct URLs: `[fact](source-url)`
  - IMPORTANT: EVERY row in the Key Facts & Observations table MUST have at least one citation link
  - For multiple facts in a single row, include a citation link for the most significant facts
  - Cite official documentation, repository pages, security databases, and other authoritative sources whenever possible
  - Include citations for:
     * ALL license details, terms, and provisions
     * ALL security vulnerabilities and patches
     * ALL maintenance metrics and observations
     * ALL dependency numbers and management approaches
     * ALL regulatory tools and frameworks
  - If information was found on a source without a public URL (e.g., local analysis), clearly state this but still provide the observation
  - ALWAYS link to primary sources rather than secondary sources when possible (e.g., GitHub repo over blog post)
  - Include SPECIFIC links to exact locations (e.g., link to specific GitHub issue page, not just GitHub home)
  - Example: Instead of just "[TensorFlow GitHub](https://github.com/tensorflow/tensorflow)", use "[TensorFlow has 58,000+ stars](https://github.com/tensorflow/tensorflow)"

- Risk Rating Format:
  - ALWAYS use star ratings only: ⭐⭐⭐⭐⭐, ⭐⭐⭐, ⭐
  - Low risk: ⭐⭐⭐⭐⭐
  - Medium risk: ⭐⭐⭐
  - High risk: ⭐

- Risk Rating Reasons:
  - Provide a concise 1-sentence explanation for EACH risk rating
  - Explicitly reference the specific criteria that determined the rating
  - For HIGH risks, clearly state what threshold was exceeded or requirement not met
  - For LOW risks, explain what positive factors led to this favorable rating
  - When rating based on ABSENCE of information, clearly state this as the reason

- Risk Level Distribution:
  - IMPORTANT: The most realistic assessment MUST include at least ONE HIGH risk item
  - Do not artificially inflate risk; base it on evidence
  - If no clear high risk is found, identify the MOST concerning factor and explain why it poses high risk
  - Absence of critical information itself can justify a high risk rating

- Emergency Issues:
  - Include ONLY if HIGH risk with immediate impact potential is EXPLICITLY supported by sources
  - Otherwise omit this section entirely
  - Always include a specific, actionable solution
  - Never speculate about emergency scenarios not directly evidenced in sources

- Format using markdown with HTML color tags for emergency section
- Limit to maximum 350 words total
- Omit any redundant explanations or theoretical discussions
</Writing Guidelines>

<Risk Rating Criteria>
For each risk factor, apply these specific criteria:

LOW RISK (⭐⭐⭐⭐⭐):
- License: Permissive (MIT, Apache 2.0, BSD) with clear terms and compatibility
- Security: No CVEs in past 24 months, robust security policy, rapid fixes (<7 days)
- Maintenance: >10 active contributors, monthly+ releases, <24hr issue response
- Dependencies: SBOM available, <20 direct dependencies, automatic updates
- Regulatory: Clear compliance documentation, complete audit trail

MEDIUM RISK (⭐⭐⭐):
- License: Moderate restrictions or unclear patent provisions
- Security: 1-3 minor CVEs (12mo), basic security policy, moderate response (7-30 days)
- Maintenance: 3-10 contributors, quarterly releases, 1-7 day issue response
- Dependencies: Partial SBOM, 20-50 direct dependencies, some transitive visibility
- Regulatory: Incomplete compliance docs, partial audit readiness

HIGH RISK (⭐):
- License: Restrictive (GPL/AGPL), incompatible terms, legal concerns
- Security: Critical/multiple CVEs, missing security policy, slow response (>30 days)
- Maintenance: <3 contributors, infrequent releases (>6mo), poor issue response
- Dependencies: No SBOM, >50 direct dependencies, vulnerable transitive deps
- Regulatory: Missing compliance docs, fails essential regulations
- IMPORTANT: Absence of critical information on any key risk factor should be rated as HIGH RISK
</Risk Rating Criteria>

<Final Check>
1. Verify EVERY row in your Key Facts & Observations table has at least one citation link
2. Confirm all relevant risk metrics for your section are addressed
3. Ensure star ratings are used correctly
4. Confirm at least ONE high-risk (⭐) item is identified
5. Ensure EVERY risk rating has a clear reason explaining the rating
6. Ensure total length is under 350 words
7. Remove any theoretical or duplicated content
8. Verify each observation has a specific control/solution
9. Double-check that NO claims are made without explicit source evidence
10. Verify that absence of information is properly acknowledged and rated accordingly
11. Do NOT include a separate Sources section - use inline links for critical facts only
12. Do NOT use numbered citations [1], [2], etc. - ONLY use inline hyperlinks
13. Ensure there are NO notes/references/sources sections at the end of your report
14. Check that EVERY required risk factor for your section has been addressed with specific metrics
</Final Check>
\end{lstlisting}

\subsubsection{Quality Assessment Instructions for Draft Sections}
\begin{lstlisting}[caption={Quality Assessment Instructions for Draft Sections. This agent evaluates the quality of a written section and generates follow-up queries if information is missing or insufficient.}, basicstyle=\ttfamily\tiny, breaklines=true, postbreak=\mbox{{\color{red}$\hookrightarrow$}\space}, inputencoding=utf8, literate={🚨}{{[ALERT]}}{7} {⭐}{{$\star$}}{1}]
You are a Chief Information Security Officer reviewing an open source risk assessment report section:

<Library input>
{topic}
</Library input>

<section topic>
{section_topic}
</section topic>

<section content>
{section}
</section content>

<task>
Rigorously evaluate whether this section meets enterprise security standards for open source risk assessment. Apply the following STRICT evaluation criteria:

1. QUANTIFICATION: Does the section provide PRECISE metrics (exact dates, counts, percentages, time periods)?
2. EVIDENCE: Is every risk claim supported by cited source evidence?
3. RISK RATING: Is each risk factor explicitly rated (Low/Medium/High) with clear justification?
4. ACTIONABILITY: Are the recommendations specific, technical, and implementable?
5. ENTERPRISE RELEVANCE: Does the assessment address governance, compliance, and security concerns at an enterprise level?

For a PASS grade, the section must meet ALL criteria above with no significant gaps.

If any criteria are not fully met, generate {number_of_follow_up_queries} targeted follow-up search queries to obtain the missing information. These queries should be highly specific and designed to retrieve quantitative data.
</task>

<format>
Call the Feedback tool and output with the following schema:

grade: Literal["pass","fail"] = Field(
    description="Evaluation result indicating whether the risk assessment meets enterprise standards ('pass') or needs revision ('fail')."
)
follow_up_queries: List[SearchQuery] = Field(
    description="List of follow-up search queries to gather missing quantitative data.",
)
</format>
\end{lstlisting}

\subsubsection{Executive Summary Generation Instructions}
\begin{lstlisting}[caption={Executive Summary Generation Instructions. This agent generates the overall executive summary of the report, synthesizing information from all completed sections.}, basicstyle=\ttfamily\tiny, breaklines=true, postbreak=\mbox{{\color{red}$\hookrightarrow$}\space}, inputencoding=utf8, literate={🚨}{{[ALERT]}}{7} {⭐}{{$\star$}}{1}]
You are a Chief Security Officer providing the EXECUTIVE SUMMARY for an open source risk assessment report.

<Library input>
{topic}
</Library input>

<Context>
{context}
</Context>

<Task>
Create a comprehensive EXECUTIVE SUMMARY as the FIRST SECTION of the report that consolidates findings from all risk domains and includes integrated risk mitigation recommendations. The executive summary must give decision makers a complete picture of the risk profile while being concise and actionable.
</Task>

<Executive Summary Format>
## Executive Summary

### Risk Score Dashboard
| Risk Domain | Rating | Key Finding | Reason for Rating | Key Control |
|-------------|--------|-------------|-------------------|-------------|
| License     | ⭐⭐⭐⭐⭐ | [Specific metric with citation link](url) | [Why this is low risk] | [Solution]  |
| Security    | ⭐⭐⭐ | [Specific metric with citation link](url) | [Why this is medium risk] | [Solution]  |
| Maintenance | ⭐⭐⭐⭐ | [Specific metric with citation link](url) | [Why this is low risk] | [Solution]  |
| Dependencies| ⭐ | [Specific metric with citation link](url) | [Why this is high risk] | [Solution]  |
| Regulatory  | ⭐⭐⭐ | [Specific metric with citation link](url) | [Why this is medium risk] | [Solution]  |
| **OVERALL** | ⭐⭐⭐ | [Overall assessment with citation link](url) | [Why this overall rating] | [Priority action] |

### 🚨 EMERGENCY ISSUES
<span style="color:red">
**[Critical Issue]**: [Most serious HIGH risk finding with citation link](url)
* **Immediate Action**: [Specific, implementable solution]
</span>

### Top Controls by Priority
1. **Immediate (0-7 days)**: [Action for HIGH risk items with citation link](url)
2. **Short-term (30 days)**: [Important technical control with citation link](url)
3. **Medium-term (90 days)**: [Important policy/legal control with citation link](url)

### Comprehensive Risk Mitigation Strategy
Based on all section findings, provide a concise but comprehensive summary of risk mitigation actions needed across all domains:

1. **Technical Controls**:
   - [Specific technical implementation or control with citation link](url)
   - [Specific technical implementation or control with citation link](url)

2. **Policy & Governance Controls**:
   - [Specific policy or governance control with citation link](url)
   - [Specific policy or governance control with citation link](url)

3. **Legal & Compliance Controls**:
   - [Specific legal or compliance control with citation link](url)
   - [Specific legal or compliance control with citation link](url)
</Executive Summary Format>

<Guidelines>
- PLACEMENT: The Executive Summary MUST be the FIRST section of the report
- SCOPE: This summary must cover ALL risk domains assessed in the detailed sections

- CITATION FORMAT AND FREQUENCY:
  - ONLY use inline markdown hyperlinks for direct URLs: `[fact](source-url)`
  - IMPORTANT: EVERY row in the Risk Score Dashboard table MUST have at least one citation link
  - EVERY recommended control in all sections MUST include a citation link to source guidance or documentation
  - Link to specific pages and resources, not just general websites
  - Include links to:
     * ALL significant vulnerabilities and findings
     * ALL tools or frameworks mentioned
     * ALL reference documentation for recommended controls
     * ALL key metrics underpinning risk assessments
  - Example: Instead of just "[TensorFlow security page](https://www.tensorflow.org/security)", use "[12 critical CVEs reported in TensorFlow since 2022](https://www.tensorflow.org/security)"
  - Focus on links to primary sources (official documentation, repository data, security databases)
  - ALWAYS verify URLs exist before including them
  - NEVER hallucinate or fabricate links
  - If uncertain about a URL\'s existence, present the fact without a link
  - Do NOT use numbered citations or separate reference lists

- RISK RATINGS: Use star ratings only:
  - ⭐⭐⭐⭐⭐ for Low risk
  - ⭐⭐⭐ for Medium risk
  - ⭐ for High risk

- HIGH RISK: MUST identify at least one HIGH risk area (⭐)
- JUSTIFICATION: For EACH risk rating, provide a clear 1-sentence reason explaining why it received that rating
- EMERGENCY ISSUES: This section should ONLY appear if truly critical issues exist
- LENGTH: Limit to 600 words maximum for readability
- FOCUS: Present only the highest priority findings from each domain
- ACTIONABILITY: Ensure every finding has a corresponding control/solution
- ORDER: Risk domains should be ordered from highest to lowest risk
- MITIGATION SECTION: Include a dedicated risk mitigation strategy section that consolidates recommendations from all sections
- CONSISTENCY CHECK: Ensure all facts and assessments are consistent across the entire executive summary
</Guidelines>
\end{lstlisting}

\subsubsection{Repository Identification Instructions for Benchmarking}
\begin{lstlisting}[caption={Repository Identification Instructions for Benchmarking. This agent identifies the GitHub repository URL for a given library name or URL.}, basicstyle=\ttfamily\tiny, breaklines=true, postbreak=\mbox{{\color{red}$\hookrightarrow$}\space}, inputencoding=utf8, literate={🚨}{{[ALERT]}}{7} {⭐}{{$\star$}}{1}]
You are a GitHub repository identifier.

<Library input>
{topic}
</Library input>

<Full Report>
{full_report}
</Full Report>

<Task>
Extract the GitHub repository owner and name from the input. The input may be:
1. A direct GitHub URL (e.g., https://github.com/owner/repo)
2. A library name that can be mapped to a GitHub repository (e.g., "TensorFlow", "React")
3. Any other open source project reference

For library names or general references, determine the most official or popular GitHub repository.

Return the repository in the format "owner/repo".
</Task>

<Format>
Call the GitHubRepo tool
</Format}
\end{lstlisting}

\subsubsection{Scorecard Analysis and Report Comparison Instructions}
\begin{lstlisting}[caption={Scorecard Analysis and Report Comparison Instructions. This agent compares the generated report against OpenSSF Scorecard results to identify overlaps and novel findings.}, basicstyle=\ttfamily\tiny, breaklines=true, postbreak=\mbox{{\color{red}$\hookrightarrow$}\space}, inputencoding=utf8, literate={🚨}{{[ALERT]}}{7} {⭐}{{$\star$}}{1}]
You are an open source security analyst specializing in the OpenSSF Scorecard.

<Scorecard Results>
{scorecard_results}
</Scorecard Results>

<Full Report>
{full_report}
</Full Report>

<Task>
Analyze the OpenSSF Scorecard results alongside the full risk assessment report to determine:

1. Model Coverage: Which OpenSSF Scorecard metrics were already covered in the full report
2. Model Seeking: Which issues were discovered by the model but not identified by Scorecard

IMPORTANT:
- EXCLUDE all scorecard checks with "?" scores from your analysis
- The denominator for coverage should be the total number of applicable checks (excluding "?" scores)
- Count each row in the scorecard results table as one check

IMPORTANT METRICS TO TRACK:
1. MODEL_COVERAGE: Number of OpenSSF Scorecard checks that were adequately addressed in the report
2. MODEL_SEEKING: Number of issues the model found that weren't explicitly mentioned in Scorecard

FORMAT IN MARKDOWN:
Instead of using dictionaries for lacks and extras, include this information as bullet points in your coverage_summary using markdown format:

**Coverage Summary:**
- Model Coverage: [Actual covered checks]/[Total applicable checks] scorecard checks addressed in report.
- Model Seeking: [Number] issues found by model but not in Scorecard.

**Checks Missing from Report:**
- **[Name of Check]**: [Explanation of what was missed]

**Issues Found Only by Model:**
- **[Name of Issue]**: [Explanation of what model found]

You MUST use the actual numeric values from your analysis for the coverage metrics. For example, if you found that 14 out of 18 checks were covered, write "Model Coverage: 14/18".
You MUST replace bracketed placeholders like '[Actual covered checks]' with the real data from your analysis.
</Task>

<Format>
Call the ScorecardAnalysis tool
</Format}
\end{lstlisting}

\subsection{Detailed Assessment Example: JAX Library Report}
\label{subsec:ExampleJAXAssessment}
To illustrate the detailed report format generated by our system, this subsection presents the complete, multi-page risk assessment report produced by \textsc{LibVulnWatch} for the JAX library. 
This report exemplifies the structure, depth of analysis, and range of risk factors (covering License, Security, Maintenance, Dependencies, and Regulatory domains) assessed for each library. Such detailed reports aim to provide actionable insights for stakeholders.
This serves as an exemplar; upon acceptance, all generated reports for the evaluated libraries will be made publicly available via our Hugging Face Space.

\includepdf[pages=-, scale=0.9, pagecommand={\thispagestyle{empty}}]{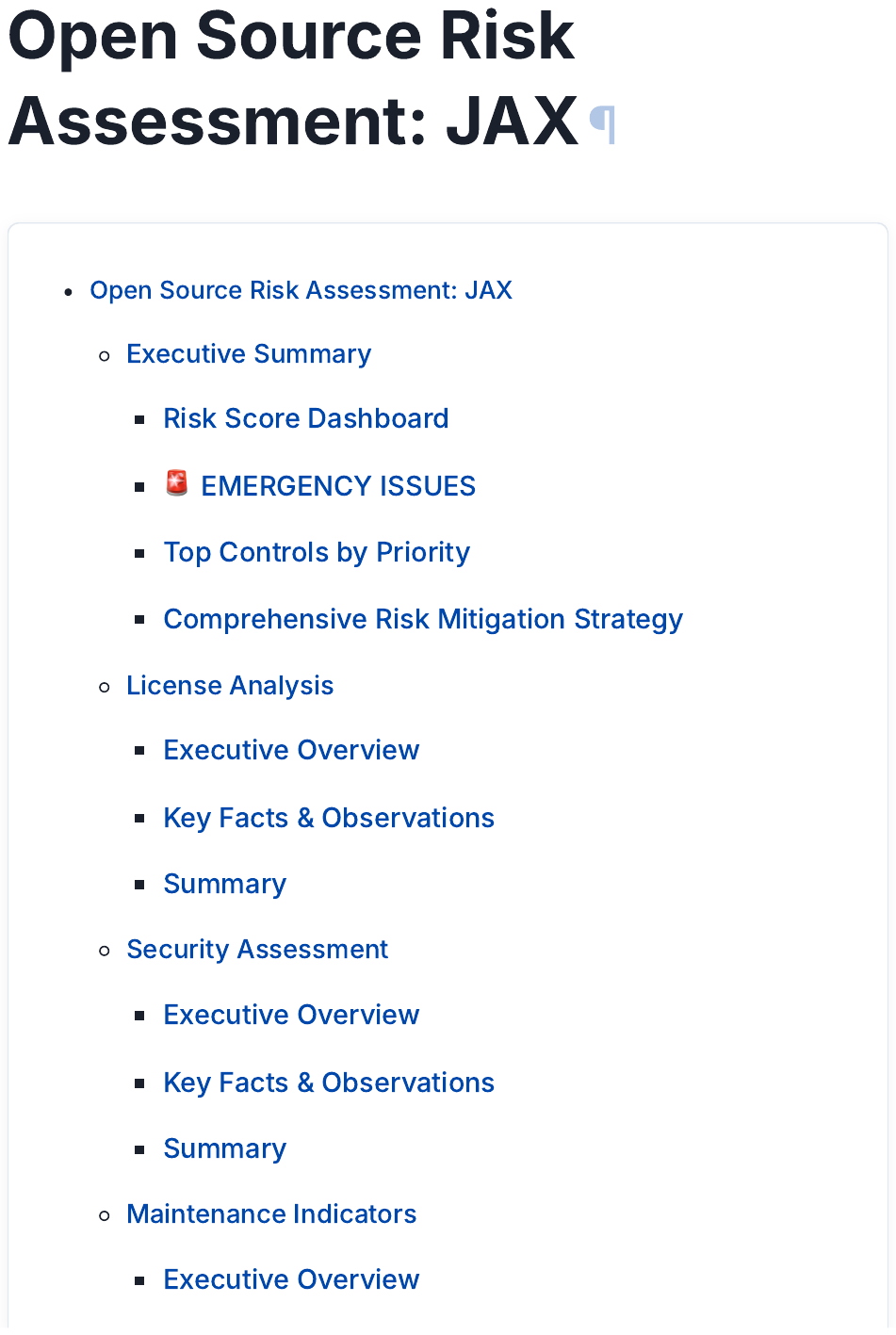}


\subsection{Example Baseline Evaluation: JAX Library Report}
\label{subsec:JAXReportEvaluation}
Following the full JAX library assessment report presented in Appendix~\ref{subsec:ExampleJAXAssessment}, this subsection provides the corresponding automated baseline evaluation. This evaluation compares \textsc{LibVulnWatch}'s findings for JAX against the OpenSSF Scorecard, detailing the alignment between the two and highlighting novel risks or deeper contextual insights uniquely identified by our agentic system. This comparative analysis is crucial for understanding the added value and specific strengths of our approach when applied to a specific library assessment.

\begin{lstlisting}[caption={JAX Library Assessment Evaluation (comparison with OpenSSF Scorecard).}, basicstyle=\ttfamily\tiny, breaklines=true, postbreak=\mbox{{\color{red}$\hookrightarrow$}\space}, inputencoding=utf8, literate={🚨}{{[ALERT]}}{7} {⭐}{{$\star$}}{1}]
**Coverage Summary:**
- Model Coverage: 11/18 scorecard checks addressed in report.
- Model Seeking: 12 issues found by model but not in Scorecard.

**Checks Missing from Report:**
- **Code-Review**: The Scorecard identified low code review approvals (3/10), but the report did not explicitly discuss code review quality or approval ratios.
- **Dangerous-Workflow**: Scorecard verified the absence of dangerous workflows but the full report did not address this workflow security aspect.
- **Dependency-Update-Tool**: Scorecard found update tools (Dependabot) used, but the full report highlighted a critical lack in dependency management and did not discuss presence of update tooling.
- **Fuzzing**: Scorecard noted no fuzzing; the report lacked any mention of fuzz or dynamic testing efforts.
- **Maintainance indicators on issue resolution**: While Scorecard gave a perfect score on Maintained, the report notes absence of issue resolution SLAs and some open issues, indicating a maintenance concern not captured in scorecard summary.
- **Packaging**: Scorecard could not assess; full report noted good automated packaging but did not discuss packaging workflow security.
- **Signed-Releases**: Scorecard could not score; the report discusses unsigned releases raising supply chain risks.

**Issues Found Only by Model:**
- **Absence of SBOM and Dependency Transparency**: The report highlights complete lack of SBOM, transitive dependency management, and vulnerability scanning as critical, absent from Scorecard findings.
- **Security Process Gaps**: Missing published security disclosure policies, patch SLAs, and CI/CD security integration not described by Scorecard.
- **Regulatory and Compliance Risks**: High regulatory risk with no GDPR, HIPAA, AI governance, or explainability support fully discussed only by model.
- **Legal Licensing Limitations for Mouse Models**: The restrictive and risky JAX Leap License for mouse models posing commercial legal risks not detected by Scorecard.
- **Dependency Vulnerability and Update Weaknesses**: While Scorecard found some update tooling, the model reveals severe vulnerability management gaps.
- **Token Permissions Excessive**: Scorecard flags token permission issues; the report does not discuss token permission risks.
- **Vulnerabilities Present**: Scorecard reports 18 existing vulnerabilities; the full report sees no recent CVEs and thus conflicts on direct vulnerability findings.

The model identified more nuanced regulatory, legal, and dependency supply chain details that the Scorecard metrics alone did not reveal, while Scorecard provided some workflow and token permissions insights not covered by the model.
\end{lstlisting}

\end{document}